\documentclass[11pt]{article}
\usepackage{graphicx}
\usepackage{amsmath,amsthm,amssymb}
\usepackage{mathptmx}
\usepackage{fullpage}
\usepackage[auth-sc]{authblk}

\newtheorem{thm}{Theorem}
\newtheorem{lem}{Lemma}
\newtheorem{prp}{Proposition}
\newtheorem*{definition}{Definition}
\newcommand{\poly}{\mathrm{poly}}

\begin{document}

\title{Fixed-point tile sets and their applications\thanks{Supported in part by ANR EMC ANR-09-BLAN-0164-01, NAFIT ANR-08-EMER-008-01, and RFBR 09-01-00709-a grants.}}
\author[1]{Bruno Durand}
\author[1,2]{Andrei Romashchenko}
\author[1,2]{Alexander Shen}

\affil[]{Laboratoire d'Informatique Fondamentale de Marseille, CNRS \& Univ. Aix--Marseille}
\affil[2]{On leave from the  Institute for Information Transmission Problems  of RAS, Moscow.}

\maketitle

\begin{abstract}
An aperiodic tile set was first constructed by R.~Berger while
proving the undecidability of the domino problem. It turned out
that aperiodic tile sets appear in many fields, ranging from
logic (the Entscheidungsproblem) to physics (quasicrystals).

We present a new construction of an aperiodic tile set that is
based on Kleene's fixed-point construction instead of geometric
arguments. This construction is similar to J.~von Neumann's
self-reproducing automata; similar ideas were also used by
P.~G\'acs in the context of error-correcting computations.

This construction is rather flexible, so it can be used in many
ways. We show how it can be used to implement substitution
rules, to construct strongly aperiodic tile sets (in which any tiling is
far from any periodic tiling), to give a new proof for the
undecidability of the domino problem and related results,
to characterize effectively closed one-dimensional subshifts in terms of two-dimensional subshifts
of finite type (an improvement of a result by M.~Hochman),
to construct a tile set that has only complex tilings, and to
construct a ``robust'' aperiodic tile set that does not have
periodic (or close to periodic) tilings even if we allow some
(sparse enough) tiling errors. For the latter, we develop a
hierarchical classification of points in random sets into
islands of different ranks.
Finally, we combine and modify our tools to prove our main
result: There exists a tile set such that all tilings have high
Kolmogorov complexity even if (sparse enough) tiling errors are
allowed.

Some of these results were included in the DLT extended
abstract~\cite{dlt} and in the ICALP extended abstract~\cite{icalp}.
\end{abstract}

\clearpage
\tableofcontents
\clearpage

\section{Introduction}

In this paper, \emph{tiles} are unit squares with colored sides.
Tiles are considered as prototypes: we may place translated
copies of the same tile into different cells of a cell paper
(rotations are not allowed). Tiles in the neighbor cells should
match (i.e., the common sides should each have the same color).

Formally speaking, we consider a finite set $C$ of
\emph{colors}. A \emph{tile} is a quadruple of colors (left,
right, top, and bottom ones), i.e., an element of $C^4$. A
\emph{tile set} is a subset $\tau\subset C^4$. A \emph{tiling}
of the plane with tiles from $\tau$ (\emph{$\tau$-tiling}) is a
mapping $U\colon \mathbb{Z}^2\to\tau$ that respects the color-matching condition.

A tiling $U$ is \emph{periodic} if it has a
\emph{period}, i.e., a nonzero vector $T\in\mathbb{Z}^2$ such
that $U(x+T)=U(x)$ for all $x\in\mathbb{Z}^2$. Otherwise, the
tiling is \emph{aperiodic}. The following classical result was
proved  in~\cite{berger}:
\begin{thm}\label{theorem1}
There exists a tile set $\tau$ such that $\tau$-tilings exist
and all of them are aperiodic.
\end{thm}
The construction from the proof of Theorem~\ref{theorem1} was used in~\cite{berger}
as the main tool to prove \emph{Berger's theorem}:
The \emph{domino problem} (to find out whether  or not a given tile set
has tilings) is undecidable.

The first tile set of Berger was rather complicated. Later, many
other constructions were suggested. Some of them are simplified
versions of  Berger's construction (\cite{robinson}; see also
the expositions in
\cite{durand-gurevich,intelligencer,levin-arxiv}). Some others
are based on polygonal tilings (including the famous Penrose and
Ammann tilings; see~\cite{grunbaum}). An ingenious construction
suggested in~\cite{kari} is based on  multiplication in a
kind of positional number system and gives a small aperiodic set
of $14$ tiles (and in~\cite{culik} an improved version with $13$
tiles is presented). Another nice construction with a short and
simple proof (based explicitly on ideas of self-similarity) was
recently proposed in~\cite{ollinger}.

In this paper, we present yet another construction of an aperiodic
tile set. It does not provide a small tile set; however, we find
it interesting for the following reasons:

\begin{itemize}
\item The existence of an aperiodic tile set becomes a
simple application of the classical construction used in Kleene's
fixed-point (recursion) theorem, in von Neumann's
self-re\-produ\-cing automata~\cite{neumann}, and, more recently, in
G\'acs' reliable cellular automata~\cite{gacs-focs,gacs}; we do
not use any geometric tricks. An aperiodic
tile set is not only an interesting result but an important tool
(e.g., this construction was invented to prove that the domino problem is
undecidable); our construction makes this tool easier to use.

\item The construction is rather general, so it is
flexible enough to achieve some additional properties of the
tile set. We illustrate this flexibility by providing new proofs for
several known results and proving new results; these new results
add robustness (resistance to sparse enough errors) to known
results about aperiodic tile sets and tile sets that have only
complex tilings.
\end{itemize}

It is unclear whether this kind of robustness can be achieved
for previously known constructions of tile sets. On the other hand, robustness properties 
appear to be important. For example, mathematical models for processes 
such as quasicrystal growth or
DNA computation should take errors into account. Note that our
model (with its independent choice of places where errors are allowed)
has no direct physical meaning; it is just a simple mathematical
model that can be used as a playground to develop tools for
estimating the consequences of tiling errors.

\medskip

The paper is organized as follows:

\begin{itemize}
\item In Section~\ref{fixed}, we
present the fixed-point construction of an aperiodic tile set
(new proof of Berger's theorem), and we illustrate the
flexibility of this construction by several examples. 

\item
In Section~\ref{substitution}, we show that any ``uniform''
substitution rule can be implemented by a tile set
(thus providing a new proof for
this rather old result).
\item
In Section~\ref{thue}, we use substitutions to show that there are strongly aperiodic tile sets (which means that any tiling is strongly aperiodic, i.e., any
shift changes at least some fixed fraction of tiles).

\item
The fixed-point construction of Section~\ref{fixed} provides a self-similar tiling: Blocks of size $n\times n$ (``macro-tiles'')
behave exactly as individual tiles, so on the next level we have
$n^2\times n^2$ blocks made of $n\times n$ macro-tiles that have
the same behavior, etc. In Section~\ref{variable}, we make some
changes in our construction that allow us to get variable zoom
factors (the numbers of tiles in macro-tiles increase as the
level increases).

Variable zoom factor tilings can be used for simulating
computations (with higher levels performing more computation steps); we
use them to give a simple proof of the undecidability of the
domino problem. The main technical difficulty in the standard proof
was to synchronize computations on different levels. In our construction 
this is not needed. We show also that other undecidability
results can be obtained in this way.

\item
This technique can be used to push the strong aperiodicity to
its limits: The distance between every tiling and every periodic configuration
(or between every tiling and its nontrivial shift) can be made
arbitrarily close to $1$, not only separated from~$0$. This is
done in Section~\ref{strongly} using an additional tool:
error-correcting codes.

\item
In~\cite{dls}, a tile set was constructed such that every tiling
has maximal Kolmogorov complexity of fragments ($\Omega(n)$ for
$n\times n$ squares); all tilings for this tile set are noncomputable (thereby implying a classical result of Hanf~\cite{hanf} and
Myers~\cite{myers} as a corollary).
The construction in~\cite{dls} was rather complicated
and was based on a classical construction of an aperiodic tile
set. In Section~\ref{complex}, we provide another proof of the same result that uses variable zoom factors. It is simpler in
some respects and can be generalized to produce robust tile sets
with complex tiling, which is our main result
(Section~\ref{robust-complex}).
\item
In Section~\ref{subshifts}, we use the same technique to give
a new proof for some results by Simpson~\cite{simpson-subshifts}
and Hochman~\cite{hochman} about effectively closed subshifts:
Every one-dimensional effectively closed  subshift can be obtained as
a projection of some two-dimensional subshift of finite
type (in an extended alphabet).
Our construction provides a solution of Problem~9.1 from~\cite{hochman}.
(Another solution, based on the classical Robinson-type construction, was
independently suggested by Aubrun and Sablik; see~\cite{sablik-aubrun}.)

\item
To prove the robustness of tile sets against sparse errors we
use a hierarchical classification of the elements of random sets
into islands of different levels (a method that goes back to
G\'acs~\cite{gacs,gray}). This method is described in
Section~\ref{random}.
In Section~\ref{islands}, we give definitions and establish some
probabilistic results about islands that are used later to prove
robustness. We show that a sparse random set on $\mathbb{Z}^2$
with probability $1$ (for Bernoulli distribution) can be represented
as a union of ``islands'' of different ranks. The higher the rank,
the bigger is the size of an island; the islands are well isolated
from each other (i.e., in some neighborhood of an island of rank $k$,
there are no other islands of rank $\ge k$).
Then, in Section~\ref{percolation}, we illustrate
these tools using standard results of percolation theory as a
model example.
In Section~\ref{bi-islands}, we modify the definition of an island
by allowing two (but not three!) islands of the same rank to be close to each other.
This more complicated definition is necessary to obtain the most
technically involved result of the paper in Section~\ref{robust-complex}
but can be skipped if the reader is interested in the other results.

\item
In Section~\ref{robust}, we use a fixed-point construction to get
an aperiodic tile set that is robust in the following sense: If
a tiling has a ``hole'' of size $n$, then this hole can be
patched by changing only an $O(n)$-size zone around it. Moreover,
we do not need for this a tiling of the entire plane. An $O(n)$ zone 
(with bigger constant in $O$ notation) around the
hole is enough.

\item In Section~\ref{robust-var}, we explain how to get
robust aperiodic tile sets  with variable zoom factors. Again,
this material is used in Section~\ref{robust-complex} only.

\item
In Section~\ref{strongly-robust}, we combine the developed techniques
to establish one of our main results: There exists a
tile set such that every tiling of the plane minus a sparse set
of random points is far from every periodic tiling.

\item
Finally,  Section~\ref{robust-complex} contains our most
technically difficult result: a robust tile set such that all
tilings, even with sparsely placed holes, have linear
complexity of fragments. To this end  we need to combine
all our techniques:
fixed-point construction with variable zoom factors,
splitting of a random set into doubled islands (we shall call them bi-islands), and
``robustification'' with  filling of holes.

\end{itemize}

\section{Fixed-point aperiodic tile set}\label{fixed}

\subsection{Macro-tiles and simulation}
        \label{macrotiles}
Fix a tile set $\tau$ and an integer $N>1$ (\emph{zoom factor}).
A \emph{macro-tile} is an $N\times N$ square tiled by
$\tau$-tiles matching each other
(i.e., a square block of $N^2$ tiles with no color
conflicts inside).  We can consider
macro-tiles as ``preassembled'' blocks of tiles; instead of tiling
the plane with individual tiles, we may use macro-tiles. To get a
correct $\tau$-tiling in this way, we need only
to ensure that neighbor
macro-tiles have matching \emph{macro-colors},
so there are no color mismatches on the borders between
macro-tiles. More formally, by macro-color
we mean a sequence of $N$ colors on the side of a macro-tile
(i.e., the right macro-color is a sequence of the right
colors of the tiles on the right edge of a macro-tile, and the same
for the left, the top, and the bottom macro-color).
Each macro-tile has four macro-colors (one for each side).
We always assume that macro-tiles are placed side to side, so the
plane is split into $N\times N$ squares by vertical and horizontal
lines.

In the following we are interested in the situation when $\tau$-tilings
can be split uniquely into macro-tiles that behave like tiles from some
other tile set $\rho$. Formally, let us define the notion of a simulation.

Let $\tau$ and $\rho$ be two tile sets, and let $N>1$ be an integer.
By \emph{simulation of $\rho$ by $\tau$ with zoom factor $N$} we mean
a mapping $S$ of $\rho$-tiles into $N\times N$ $\tau$-macro-tiles
such that the following properties hold:
\begin{itemize}
\item $S$ is injective (i.e., different tiles are mapped into different
macro-tiles).
\item Two tiles $r_1$ and $r_2$ match if and only if their images $S(r_1)$ and
$S(r_2)$ match. This means that the right color of $r_1$ equals the
left color of $r_2$ if and only if the right macro-color of $S(r_1)$ equals
the left macro-color of $S(r_2)$, and the same is true in the vertical
direction.
\item Every $\tau$-tiling can be split by vertical and horizontal
lines into $N\times N$ macro-tiles that belong to the range of $S$, and
such a splitting in unique.
\end{itemize}

The second condition guarantees that every $\rho$-tiling can be
transformed into a $\tau$-tiling by replacing each tile $r\in\rho$ by
its image, macro-tile $S(r)$.  Taking into account other conditions, we
conclude that
every $\tau$-tiling can be obtained in this way, and the positions
of grid lines as well as the corresponding $\rho$-tiles can be reconstructed uniquely.

\textbf{Example 1} (negative). Assume that $\tau$ consists of one tile with four
white sides.
Fix some $N>1$. There exists a single macro-tile of size $N\times N$.
Does this mean that $\tau$ simulates itself (when its only tile is mapped
to the only macro-tile)? No. The first and second conditions are true,
but the third one is false: The placement of cutting lines is not unique.

\begin{figure}[h]
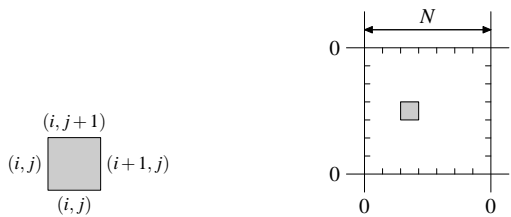

\begin{center}
\includegraphics[scale=0.7]{fpt-eng-1.mps}\hspace*{2cm}
\includegraphics[scale=0.8]{fpt-eng-2.mps}
\end{center}
\caption{Tiles and macro-tiles for Example~2.}
\label{fpt.1-2.mps}
\end{figure}

\textbf{Example 2} (positive). In this example $\rho$ consists
of one tile with all white sides. The tile set
$\tau$ consists of $N^2$ tiles indexed by pairs $(i,j)$ of
integers modulo~$N$. A tile from $\tau$ has colors on its sides
as shown on Fig.~\ref{fpt.1-2.mps} (each color is a pair of integers modulo $N$,
so set $C$ of all colors consists of $N^2$ elements). The simulation maps the single
$\rho$-tile to a
macro-tile that has
colors $(0,0),\ldots,(0,N-1)$ and $(0,0),\ldots,(N-1,0)$ on its
vertical and horizontal borders, respectively (see Fig.~\ref{fpt.1-2.mps}).

\begin{definition}
A \emph{self-similar} tile set is a tile set that
simulates itself.
\end{definition}

The idea of self-similarity is used (more or less explicitly) in
most constructions of aperiodic tile sets (but \cite{kari,culik} are
exceptions). However, not all of these constructions provide literally
self-similar tile sets in our sense.

It is easy to see that self-similarity guarantees aperiodicity.
\begin{prp}
        \label{selfsimilar-aperiodic}
A self-similar tile set $\tau$ may have only aperiodic tilings.
\end{prp}

\begin{proof}
Let $S$ be a simulation of $\tau$ by itself with zoom factor $N$.
By definition, every $\tau$-tiling $U$ can be uniquely split into $N\times
N$ macro-tiles from the range of $S$. So every period $T$ of $U$ is a
multiple of $N$ (since the $T$-shift of a cut is also a cut, the shift should respect
borders between macro-tiles).
Replacing each macro-tile by its $S$-preimage, we get a $\tau$-tiling
that has period $T/N$.
Therefore, $T/N$ is again a multiple of $N$. Iterating
this argument, we conclude that $T$ is divisible by $N^k$ for
every $k$, so $T$.\end{proof}

Note also that every self-similar tile set has arbitrarily large
finite tilings. Starting with some tile, we apply $S$ iteratively
and get a big tiled square. The standard compactness argument
guarantees the existence of a tiling of the entire plane. Therefore, to
prove the existence of aperiodic tile sets it is enough to
construct a self-similar tile set.
\begin{thm}
        \label{selfsimilar}
There exists a self-similar tile set $\tau$.
\end{thm}
Theorem~\ref{selfsimilar} was explicitly formulated and proven by
Ollinger~\cite{ollinger}; in his proof a self-similar tile set (consisting of $104$ tiles)
is constructed explicitly.  This
tile set is then used to implement substitution rules
(cf. Theorem~\ref{substitution-tiling} below).
Another example of a self-similar tile set (with many more tiles) is
given in~\cite{intelligencer}. (Note that the definition
of self-similarity used in~\cite{intelligencer} is a bit stronger.)

\medskip

We prefer a less specific and more flexible argument
based on the fixed-point idea. Our proof works for a vast class of tile sets
(though we cannot provide explicitly an aperiodic tile set of a reasonably
small size).
The rest of this section is devoted to our proof of Theorem~\ref{selfsimilar}.
Before we prove this
result,  we explain a few techniques used in our construction and show
how to
simulate a given tile set by embedding computations.

\subsection{Simulating a tile set}
        \label{simulating}

Let us start with some informal discussion. Assume that we have
a tile set $\rho$ whose colors are $k$-bit strings
($C=\{0,1\}^k$) and the set of tiles $\rho\subset C^4$ is
presented as a predicate $R(c_1,c_2,c_3,c_4)$ with four $k$-bit
arguments. Assume that we have some Turing machine $\mathcal{R}$
that computes $R$. Let us show how to simulate $\rho$ using some
other tile set $\tau$.

This construction extends Example~2, but it simulates a tile set
$\rho$ that contains not a single tile but many tiles. We keep
the coordinate system modulo $N$ embedded into tiles of $\tau$;
these coordinates guarantee that all $\tau$-tilings can be
uniquely split into blocks of size $N\times N$ and every tile
``knows'' its position in the block (as in Example~2). In
addition to the coordinate system, now each tile in $\tau$
carries supplementary colors (from a finite set specified below)
on its sides. These colors form a new ``layer'' which is superimposed
with the old one; i.e., the set of colors is now a Cartesian
product of the old one and the set of colors used in this layer.

On the border of a macro-tile (i.e., when one of the coordinates
is zero) only two supplementary colors (say, $0$ and $1$) are
allowed. So the macro-color encodes a string of $N$ bits (where
$N$ is the size of macro-tiles). We assume that $N$ is
much bigger than $k$ and let
$k$ bits in the middle of macro-tile sides represent colors from
$C$. All other bits on the sides are zeros. (This is a
restriction on tiles: Each tile ``knows'' its coordinates so it
also knows whether nonzero supplementary colors are allowed.)

\begin{figure}[h]
        $$
\includegraphics[scale=1.0]{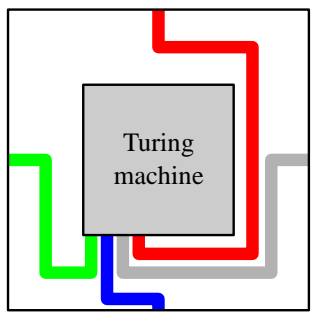}
        $$
\caption{Wires and processing zones;
          wires appear quite narrow since $N\gg k$.}
\label{fpt.3.mps}
\end{figure}

Now we need additional restrictions on tiles in $\tau$ that
guarantee that macro-colors on the sides of each macro-tile
satisfy relation $R$. To achieve this, we ensure that bits
from the macro-tile sides are transferred to the central part of
the tile where the checking computation of $\mathcal{R}$ is
simulated (Fig.~\ref{fpt.3.mps}).

For that we need to fix which tiles in a macro-tile form
``wires'' (this can be done in any reasonable way; we assume
that wires do not cross each other) and then require that each
of these tiles carries equal bits on two sides (so some bit
propagates along the entire wire); again this is easy to arrange
since each tile knows its coordinates.

Then, we check $R$ by a local rule that guarantees that the
central part of a macro-tile represents a time-space diagram of
$\mathcal{R}$'s computation (with the tape being horizontal, and time increasing
upward). This is done in a standard way:
The time-space diagram (tableau) of
a Turing machine computation can be described by local
rules, and these rules can be embedded into a tile set\footnote{%
     Speaking about local rules, we mean that one can check the correctness
     of the time-space diagram by looking through a $O(1)$-size window; in the standard
     representation width $3$ and height $2$ is enough. However, our definition
     of a tile set is even more local:  We compare colors on matching sides only. It is easy
     to see that we can still simulate any local rules by tiles. Each tile keeps
     the contents of the corresponding window, and colors are used to ensure
     that overlapping windows are consistent.\label{local-rules-footnote}%
}
(see details in, e.g., \cite{durand-gurevich,grunbaum}).
We require that computation
terminates in an accepting state; if not, the tiling cannot be
formed.

To make this construction work, the size of the macro-tile ($N$)
should be large enough; we need enough space for $k$ bits to
propagate and enough time and space (= height and width) for all
accepting computations of $\mathcal{R}$ to terminate.

In this construction the number of supplementary colors depends
on the machine $\mathcal{R}$ (the more states it has, the more
colors are needed in the computation zone). To avoid this
dependency, we replace $\mathcal{R}$ by a fixed universal Turing
machine~$\mathcal{U}$ that runs a \emph{program}
simulating~$\mathcal{R}$. Let us agree that the tape of the
universal Turing machine has an additional read-only layer. Each
cell carries a bit that is not
\begin{figure}[h]
 \begin{center}
        $$
\includegraphics[scale=1.0]{fpt-eng-4.mps}
        $$
 \end{center}
 \caption{Checking tiles with a universal Turing machine.}
\label{fpt.4.mps}
\end{figure}
changed during the computation; these bits are used as a program
for the universal machine~$\mathcal{U}$. (We may assume that the program bits occupy some part
of the reserved read-only layer, e.g., the leftmost bits on this layer; see Fig.~\ref{fpt.4.mps}.) In terms
of our simulation, the columns of the computation zone carry
unchanged bits (considered as a program for~$U$), and the tile
set restrictions guarantee that the central zone represents the
\emph{record} (time-space diagram) of an accepting
computation of~$\mathcal{U}$ (with this program).
In this way, we get a tile set $\tau$ that simulates $\rho$ with
zoom factor $N$ using $O(N^2)$ tiles. (Again we need $N$ to be
large enough, but the constant in $O(N^2)$ does not depend
on~$N$.)

\subsection{Simulating itself}
        \label{simulating-itself}
We know how to simulate a given tile set $\rho$ (represented as
a program for the universal Turing machine) by another tile set $\tau$ with
a large enough zoom factor $N$. Now we want $\tau$ to be
identical to $\rho$ in which case Proposition~\ref{selfsimilar-aperiodic}
guarantees aperiodicity). For this we use a construction that
follows the proof of Kleene's recursion (fixed-point) theorem.

We cannot refer here to the \emph{statement} of the theorem;
we need to recall its proof and adapt it to our framework.
    Kleene's theorem~\cite{kleene} says that for every computable
    transformation $\pi$ of programs one can find a program $p$
    such that $p$ and $\pi(p)$
    are equivalent, i.e., produce the same output. (For simplicity we consider
    programs with no input, but this restriction does not really matter.) In other
    words, there is no guaranteed way to transform a given program $p$ into
    some other program $\pi(p)$ that produces different output.
    As a sketch of the proof, first we note that the statement is
    language-independent since we may use
    translations in both directions before and after $\pi$. Therefore, without
    loss of generality, we
    may assume that the programming language has some
    special properties. First, we assume that it has a function
    \texttt{GetText()} that returns the text of the program (or a pointer
    to a memory address where the program text is kept).
    Second, we assume that the language contains
    an interpreter function \texttt{Execute(string s)}
    that interprets the content of its string argument \texttt{s} as a program
    written in the same language. It is not difficult to develop
    such a language and write an interpreter for it.
    Indeed, the interpreter can access the program text anyway,
    so it can copy the text into some string variable. The interpreter also can
    recursively call itself with another program as an argument when it
    sees the \texttt{Execute} call. If our language has these
    properties, it is easy to construct the fixed point for $\pi$: Just
    take the program \texttt{Execute(}$\pi$\texttt{(GetText()))}.

    This theorem shows that a kind of self-reference, in which we
    write the program as if its full text is already given to us, is
    still acceptable. A classical example is a program that prints
    its own text. The proof shows a way how to do this by using
    a computation model where the immutable text of the program
    is accessible to it.

Constructing a self-similar tiling, we have the same kind
of problems. We have already seen how to construct a tile set $\tau$
that simulates a given tile set $\rho$. [Counterpart: It is easy to write
a program that prints any given text.] What we need is to construct
a tile set that simulates \emph{itself}. [Counterpart: What we need is to
write a program that prints \emph{its own} text.]

Let us look again at our construction that transforms the
description of $\rho$ (a Turing machine that computes
the corresponding predicate) into a tile set $\tau$ that simulates $\rho$.
Note that most rules of $\tau$ do not depend on the program for
$\mathcal{R}$, dealing with information transfer along the
wires, the vertical propagation of unchanged program bits, and
the space-time diagram for the universal Turing machine in the computation
zone. Making these rules a part of $\rho$'s definition (by
letting $k=2\log N + O(1)$ and encoding $O(N^2)$ colors by $2\log
N+O(1)$ bits), we get a program that checks that macro-tiles
behave like $\tau$-tiles in this respect. Macro-tiles of the second
level (``macro-macro-tiles'') made of them would have the correct
structure, wires that transmit bits to the computation zone, and
even the record of some computation in this zone, but this
computation could have an arbitrary program. Therefore, at the third level all the
structure is lost.

What do we need to add to our construction to close the
circle and get self-simulation?
The only remaining part of the rules for $\tau$ (not implemented yet
at the level of macro-tiles) is the hard-wired
program. We need to ensure that macro-tiles carry the same
program as $\tau$-tiles do. For that our program (for the
universal Turing machine) needs to access the bits of its own text. As we
have discussed, this
self-referential action is in fact quite legal: The program is
written on the tape, and the machine can read it. The program
checks that if a macro-tile belongs to the first line of the
computation zone, this macro-tile carries the correct bit of the
program.

How should we choose $N$ (hard-wired in the program)? We need it
to be large enough so the computation described above (which deals
with $O(\log N)$ bits) can fit in the computation zone.
Note that the computation never deals with the list of
tiles in $\tau$ or a truth table of the corresponding $4$-ary relation on bit strings; all these objects are represented by programs that
describe them. The computation needs to check simple things only:
that numbers in the $0,\ldots ,N-1$ range on four sides are consistent
with each other, that rules for wires and computation time-space
diagram are observed, that program bits on the next level coincide
with actual program bits, etc.
All these computations are rather simple. They are polynomial
in the input size, which is $O(\log N)$), so for large $N$ they easily fit in
$\Omega(N)$ available time and space.

This finishes the construction of a self-similar aperiodic tile
set.

\textbf{Remark}. \label{polymorphism}
Let us also make a remark that will be useful later. We defined
a tile set as a subset of $C^4$, where $C$ is a set of colors. Using this
definition, we do not allow different tiles to have the same colors on their
sides. The only information carried by the tile is kept on its sides. However,
sometimes a more general definition is preferable. We can define a tile set
as a finite set $T$ together with a mapping of $T$ into $C^4$. Elements
of $T$ are tiles, and the mapping tells us for each tile which colors it has on
its four sides.

One can easily extend the notions of macro-tiles and simulation to
this case. In fact, macro-tiles are well suited to this definition since they
already may carry  information that is not reflected in the side macro-colors.
The construction of a self-similar tile set also can be adapted. For example, we
can construct a self-similar tile set where each tile carries an auxiliary
bit, i.e., exists in two copies having the same side colors. Since the tile
set is self-similar, every macro-tile at every level of the hierarchy also
carries one auxiliary bit, and the bits at different levels and in different macro-tiles
are unrelated to each other.
Note that the total density of information contained in a tiling is still
finite, since the density of information contained in auxiliary bits assigned to high-level
macro-tiles decreases with level as a geometric sequence.

\section{Implementing substitution rules}
\label{substitution}

The construction of a self-similar tiling is rather flexible and
can be easily augmented to get a self-similar tiling with
additional properties. Our first illustration is the simulation
of substitution rules.

Let $A$ be some finite alphabet and $m>1$ be an integer. A
\emph{substitution rule} is a mapping $s\colon A\to A^{m\times
m}$. This mapping can be naturally extended to
$A$-configurations. By \emph{$A$-configuration} we mean an
integer lattice filled with $A$-letters, i.e., a mapping
$\mathbb{Z}^2\to A$ considered modulo translations. A
substitution rule $s$ applied to a configuration $X$ produces
another configuration $s(X)$ where each letter $a\in A$ is
replaced by an $m\times m$ matrix $s(a)$.

We say that a configuration $X$ is \emph{compatible} with substitution rule
$s$ if there exists an infinite sequence
        $$
\cdots \stackrel{s}{\to} X_{3}
\stackrel{s}{\to} X_{2}
\stackrel{s}{\to} X_{1}
\stackrel{s}{\to} X,
        $$
where $X_i$ are some configurations.
This definition was proposed in \cite{ollinger}. The classical definition
(used, in particular, in \cite{mozes}) is slightly different:  Configuration
$X  :  \mathbb{Z}^2\to A$
is said to be compatible with a substitution rule $s$ if every finite part of $X$
occurs inside of some $s^{(n)}(a)$ (for some $n\in\mathbb{N}$ and some
$a\in A$). We prefer the first approach since it looks more natural in the
context of tilings. However, all our results can be reformulated and proven
(with some technical efforts) for the other version
of the definition; we do not go into details here.

\textbf{Example 3}. Let $A=\{0,1\}$,
        $$
s(0)=(\begin{smallmatrix} 0 & 1 \\ 1 & 0
\end{smallmatrix}),\quad s(1)=(\begin{smallmatrix} 0 & 1 \\ 1 &
0 \end{smallmatrix}).
        $$
It is easy to see that the
only configuration compatible with~$s$ is the chess-board
coloring where zeros and ones alternate horizontally and vertically.

\begin{figure}[h]
\begin{center}
\includegraphics[scale=1.0]{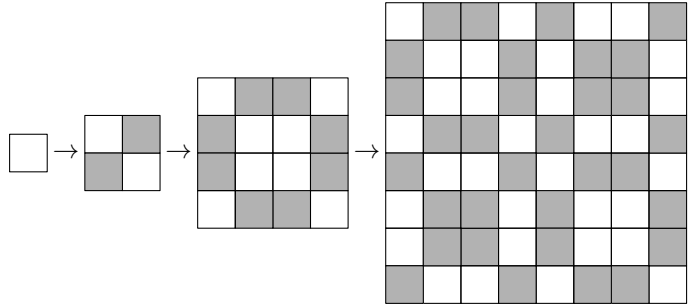}
\end{center}
\caption{Three steps of Thue--Morse substitution.}
\label{fpt-eng.45}
\end{figure}

\textbf{Example 4} (Fig.~\ref{fpt-eng.45}). Let $A=\{0,1\}$,
        $$
s(0)=(\begin{smallmatrix} 0 & 1 \\ 1 & 0 \end{smallmatrix}),\quad
s(1)=(\begin{smallmatrix} 1 & 0 \\ 0 & 1 \end{smallmatrix}).
        $$
One can check that all configurations that are compatible with
this substitution rule (called \emph{Thue--Morse
configurations} in the following) are aperiodic. 
(In Section~\ref{thue} we will prove a  stronger version of this fact.)
 One may note, for example,
 that every configuration compatible with this substitution
rule can be uniquely decomposed into  disjoint $2\times 2$ blocks
$(\begin{smallmatrix} 0 & 1 \\ 1 & 0 \end{smallmatrix})$
and $(\begin{smallmatrix} 1 & 0 \\ 0 & 1 \end{smallmatrix})$ by vertical and horizontal lines;
since neighbor cells of the same color should be separated by one of those lines,
the position of the lines is unique.
Then, we can apply the argument from Proposition~\ref{selfsimilar-aperiodic}
(with $N=2$).

The following theorem goes back to Mozes~\cite{mozes}. It says
that every substitution rule can be enforced by a tile set.

\begin{thm}
       \label{substitution-tiling}
Let $A$ be an alphabet and let $s$ be a substitution rule
over~$A$. Then, there exist a tile set $\tau$ and a mapping
$e\colon\tau\to A$ such that

\textup{(a)}~the $e$-image of any $\tau$-tiling is an $A$-configuration
compatible with $s$\textup;

\textup{(b)}~every $A$-configuration compatible with $s$ can be
obtained in this way.
\end{thm}
A nice proof of this result for $2\times 2$ substitutions is given
in~\cite{ollinger}, where an explicit construction of a tile set $\tau$ for
every substitution rule  $s$ is
provided. We prove this theorem using our fixed-point argument.
In this way we avoid  the boring technical details; but
the tile sets that can be extracted from our proof contain 
a huge number of tiles.

\begin{proof}. Let us modify the construction of the tile set
$\tau$ (with zoom factor $N$) by taking $s$ into account.
First consider a very special case when

\begin{itemize}
\item  the substitution rule maps each $A$-letter into an
$N\times N$ matrix (i.e., $m=N$) and

\item the substitution rule is easy to compute: Given a letter
$u\in A$ and $(i,j)$, we can compute the $(i,j)$-th letter of
$s(u)$ in a time much less than $N$. 

\end{itemize}

In this case we proceed as follows. In our basic construction
every tile knows its coordinates in the macro-tile and some
additional information needed to arrange ``wires'' and simulate
calculations of the universal Turing machine.\footnote{We use this anthropomorphic terminology
in the hope it makes the proof more intuitive. By saying ``each tile knows its coordinates,''
we mean that the tile set is split into  $N^2$ disjoint groups; each group corresponds to
tiles that appear in one of  $N^2$ positions in the macro-tiles. The correct
positioning of the tiles is ensured (as we have seen) by
side colors. The self-similarity guarantees that the same
is true for macro-tiles, where the group (i.e., the
coordinates in a macro-tile of the next level)
is determined by the content of the computation zone
and corresponding bits (macro-colors) on the sides.}
Now, in addition to this basic
structure, each tile keeps two letters of $A$. The first is the
label of a tile itself, and the second is the label of the
$N\times N$ macro-tile it belongs to. This means that we keep
additional $2\log |A|$ bits in each tile, i.e., multiply the
number of tiles by $|A|^2$.  It remains to explain how the local
rules work. We add two requirements:

\begin{description}

\item{(i)}
\emph{The second letter is the same for neighbor tiles \textup(unless
they are separated by a border of some $N\times N$ macro-tile\textup)}.
This constraint can be easily enforced by colors on sides of tiles.
We multiply  the number of colors in our basic construction
by $|A|$; now each  color of the new construction is
a pair: its first component is a color from the basic construction and
its second component is a letter of $A$. The
second component of the new color guarantees that every two
neighbor tiles keep the same ``father'' letter
(unless these tiles are separated by a border and do not belong
to the same father macro-tile, in which case we do not exhibit the letter to those
borders).

\item{(ii)} \emph{The first letter in a tile is determined by the second
letter and the coordinates of the tile inside the macro-tile,
according to the substitution rule}. Indeed, each tile
``knows'' its coordinates in a macro-tile. Therefore, its first letter
must  appear in $s(\mbox{second letter})$ at the corresponding
position. We do not need to extend the set of  colors to enforce this
property. This requirement is only a restriction
on tiles. It explains which combinations
$$\langle\mbox{coordinates in the father macro-tile},
\mbox{first letter},
\mbox{second letter}\rangle$$
can be combined in one tile of our tile set.\footnote{A natural question arises:
What does it mean to add a letter that is determined by other information? Adding a letter
means that we create $|A|$ copies of the same tile (with different letters); but then,
the restriction prohibits all of them except one, so is there any change at all?
In fact, the actual change is
occurring on higher levels: We want the macro-tiles to have both letters written on
the tape as binary strings (in some prearranged places). This is important for
checking consistency between levels.}

\end{description}

We want the new tile set to be self-similar. Therefore, we should guarantee
that the  requirements (i) and (ii) hold also for macro-tiles.
Fortunately,
both requirements are easy to integrate in our basic self-referential
construction. In each macro-tile, two letters of $A$ are encoded by
strings of bits in some specially reserved locations on the tape of the
Turing machine (simulated in the computation zone of this macro-tile).
Requirement (i) is enforced by adding extra $\log |A|$ bits
to macro-colors; to achieve (ii), a
macro-tile should check that its first letter appears in
$s(\mbox{second letter})$ at the required position.
This is possible  when $s$ is easy to compute. (Knowing the coordinates
and the second letter, the program computes the required value
of the first letter and then compares it with the actual value.)

Requirements (i) and (ii) ensure  that if we take first letters from $A$
assigned to each tile, we get an $A$-configuration that is an
$s$-image of some other configuration. Also (because of
self-similarity) we have the same property on the level of macro-tiles.
But this is not enough. We need to guarantee that the first letter
on the level of macro-tiles is identical to the second letter on
the level of tiles. This is also achievable. The first letter of
a  macro-tile is encoded by bits in its computation zone, and we can
require that those bits match the second letter of the
tiles at that place. (Recall that the second letter is the same across the
tiles that constitute one macro-tile; note also that each tile ``knows''
its coordinates and can determine whether it is in the zone for the first letter in the macro-tile
and which bit should be there.)  By self-similarity,
the same arguments work for macro-tiles of all levels.
It is easy to see that now the tile set $\tau$ has the required
properties (each tiling projects into a configuration compatible
with~$s$ and vice versa).

However, this construction assumes that $N$ (the zoom factor) is
equal to the matrix size in the substitution rule, which is
usually not the case. In fact, usually the value of $m$ (a parameter of the substitution rule)
is fixed in advance,
and we have to choose $N$, which needs to be large enough. To overcome this difficulty,
we let $N$ be equal to $m^k$ for some $k$, and we use the
substitution rule $s^k$, i.e., the $k$th iteration of~$s$ (a
configuration is compatible with $s^k$ if and only if it is
compatible with $s$). Now we do not need $s$ to be easily
computable: For  every $s$, if $k$ is  large enough, the computation of $s^k$ will fit into
the available space  (exponential in $k$). \end{proof}

\section{The Thue--Morse lemma and strongly aperiodic tile sets}
\label{thue}

Let $\alpha>0$ be a real number. We say that a configuration
$U\colon\mathbb{Z}^2\to A$ is \emph{$\alpha$-aperiodic} if for
every nonzero vector $T\in\mathbb{Z}^2$ there exists $N$ such
that in every square whose side is at least $N$ the fraction of
points $x$ such that $U(x)\ne U(x+T)$ exceeds~$\alpha$.

\medskip
\textbf{Remark}. If $U$ is $\alpha$-aperiodic, then the Besicovitch
distance between $U$ and any periodic pattern is at least
$\alpha/2$. (The Besicovitch distance between two configurations
is defined as $\limsup_N
d_N$, where $d_N$ is the fraction of points where two configurations
differ in the $N\times N$ centered square. It is easy to see
that the distance does not depend on the choice of the center point.)

\begin{thm}
        \label{aperiodic}
There exists a tile set $\tau$ such that $\tau$-tilings exist
and every $\tau$-tiling is $\alpha$-aperiodic for every
$\alpha<1/4$.
\end{thm}

\begin{proof}
The proof is obtained by applying Theorem~\ref{substitution-tiling}
to the Thue--Morse substitution rule $T$ (Example 4). Let $C$ be a
configuration compatible with $T$. We have to show that $C$ is
$\alpha$-aperiodic for every $\alpha<1/4$. It is enough
to prove  that every configuration compatible with
the Thue--Morse substitution rule is $\alpha$-aperiodic.

Informally, we can reduce the statement to the one-dimensional case,
since Thue--Morse substitution is an \texttt{xor}-combination of two one-dimensional
substitutions. Here are the details.

Consider a one-dimensional substitution system with two rules,
$0 \to 01$ and $1\to 10$. Applying these rules to $0$ and $1$,
we get
        \begin{align*}
&0\to 01\to 0110\to 01101001 \to \ldots,\\
&1\to 10\to 1001\to 10010110 \to \ldots
        \end{align*}
Let $a_n$ and $b_n$ be the $n$th terms in these sequences
($a_0=0$, $a_1=01$, \ldots, $b_0=1$,
$b_1=10$, etc.); it is easy to see that $a_{n+1}=a_nb_n$ and
$b_{n+1}=b_na_n$.

For some $n$ we consider the \texttt{xor}-combination of
these strings, where the $(i,j)$-th bit is \texttt{xor} of the $i$th bit in the
first string and the $j$th bit in the second string. Since $b_n$ is a
bitwise negation of $a_n$, we get only two different combinations
(one obtained from two copies of $a_n$ or two copies of $b_n$, and
the other obtained from different strings), which are bitwise opposite.
It is easy to see (e.g., by induction) that these two square patterns
are images of $0$ and $1$ after $n$ steps of two-dimensional
Thue--Morse substitution.

To
prove the statement for aperiodicity of the Thue--Morse configuration,
we start with an estimate for
(one-dimensional) aperiodicity of $a_n$ and $b_n$:

\begin{lem} [folklore] For any integer $u>0$ and for any $n$
such that $u\le |a_n|/4$ the shift by $u$ steps to the right
changes at least $|a_n|/4$ positions in $a_n$ and leaves
unchanged at least $|a_n|/4$ positions. \textup(Formally, in the range
$1,\ldots, (2^n-u)$ there exist at least $(1/4)\cdot 2^n$ positions $i$
such that the $i$th and  the $(i+u)$-th bits in $a_n$ coincide and at
least $(1/4)2^n$ positions where these bits differ.\textup) \label{lemma-folklor}
\end{lem}
\begin{proof} String $a_n$ can be represented as
        $
a b b a b a a b
        $,
where $a=a_{n-3}$ and $b=b_{n-3}$. One may assume without loss
of generality that $u\ge |a|$ (or otherwise we apply Lemma~\ref{lemma-folklor}
separately to the two halves of $a_n$). Note that $ba$ appears
in the sequence twice: 
once preceded by a copy of $a$ and once preceded by a copy of $b$. 
Since these copies have opposite bits, the shifted bits match in
one of the cases and do not match in the other one.
The same is true for $ab$, which appears
preceded both by $a$ and $b$. \end{proof}

Now consider a large $N\times N$ square in a two-dimensional
Thue--Morse configuration and some shift vector $T$. We assume
that $N$ is much bigger than components of $T$ (since we are interested
in the limit behavior as $N\to\infty$). Moreover, we may assume that some
power of $2$ (let us call it $m$)
is small compared to $N$ and large compared to $T$. Then, the
$N\times N$ square consists of a large number of $m\times m$
Thue--Morse blocks and some boundary part (which can be ignored
by changing $\alpha$ slightly).  Then, we can consider each $m\times m$
block separately to estimate the fraction of positions that are changed
by the $T$-shift. If $T$ is horizontal or vertical, we can use the
statement of the lemma directly: At least 1/4 of all positions are
changed. If not (i.e., if the shift has two nonzero components), we are interested
in the
probability of some event that is an \texttt{xor} combination of two \emph{independent}
events with probabilities in the interval $(1/4,3/4)$. It is easy to check that such
an event also has probability in $(1/4,3/4)$ (in fact, even in
$(3/8,5/8)$, but we do not need this stronger bound).

Theorem~\ref{aperiodic} is proved.\end{proof}

In fact, the bound $1/4$ can be replaced by $1/3$ if we use a more
professional analysis of the Thue--Morse sequence (see,
e.g.,~\cite{thue-morse-analysis}). However, if we want to get the
strongest result of this form and make the bound close to~$1$, this
substitution rule does not work. We can use some
other rule (in a bigger alphabet) as Pritykin and Ulyashkina
have shown~\cite{pri-ul}, but we prefer to give another
construction with variable zoom factors (see Section~\ref{strongly}).

\section{Variable zoom factor}
\label{variable}

The fixed-point construction of an aperiodic tile set is flexible
enough and can be used in other contexts. For example, the
``zoom factor'' $N$ could depend on the level. This means
that instead of one tile set $\tau$ we have a sequence of tile
sets $\tau_0,\tau_1,\tau_2,\ldots$, and instead of one zoom
factor $N$ we have a sequence of zoom factors $N_0,N_1,\ldots$.  The tile set $\tau_0$ simulates $\tau_1$ with zoom factor $N_0$,
the tile set $\tau_1$ simulates $\tau_2$ with zoom factor $N_1$,
etc.

In other words, $\tau_0$-tilings can be uniquely split (by
horizontal and vertical lines) into $N_0\times N_0$ macro-tiles
from some list, and the macro-tiles in this list are in one-to-one
correspondence (which respects matching rules) with $\tau_1$. So
$\tau_0$-tilings are obtained from $\tau_1$-tilings by replacing
each $\tau_1$-tile by the corresponding $\tau_0$-macro-tile, and
each $\tau_0$-tiling has a unique reconstruction.

Further, every $\tau_1$-tiling can be split into macro-tiles
of size $N_1\times N_1$
that correspond to $\tau_2$-tiles. So after two steps of zooming out,
every $\tau_0$-tiling looks like a $\tau_2$-tiling; only a closer look
reveals that each $\tau_2$-tile is in fact a $\tau_1$-macro-tile of size
$N_1\times N_1$, and an even closer look is needed to realize that every
$\tau_1$-tile in these macro-tiles is in fact a $\tau_0$-macro-tile
of size $N_0\times N_0$.

For such a $\tau_0$-tiling we can say that it consists of level $1$ macro-tiles
of size $N_0\times N_0$ (isomorphic to $\tau_1$); at the same time it consists
of level $2$ macro-tiles of size $N_0N_1\times N_0N_1$ (isomorphic to $\tau_2$), 
etc.

This is what we want to achieve (together with other things
needed to get the tile set with desired properties). How do we achieve this?
Each macro-tile
should ``know'' its level: A macro-tile that simulates a $\tau_k$-tile
and is made of $\tau_{k-1}$-tiles,
should have  $k$ in some
place on the tape of  the Turing machine simulated in this macro-tile. To make
this information consistent between neighbors, $k$ is exhibited as a
part of the macro-colors at all four sides. The value of $k$ is used for the
computations. Macro-colors on the sides of a macro-tile
encode the coordinates of this macro-tile inside its father, and the
computation should check that they are consistent modulo $N_{k}$
(i.e., the $x$ coordinate on the right side should be equal to the $x$ coordinate
on the left side plus $1$ modulo $N_{k}$, etc.). This means that $N_{k}$
should be computable from $k$; moreover, it should be computable fast
enough to fit into the computation zone (which carries only $\Theta(N_{k-1})$
steps of computation). After $N_{k}$ is computed, there should be enough
time to perform the arithmetic operations modulo $N_{k}$, and
so on.

Let us look at these restrictions more closely.
We need to keep both $k$ and the coordinates (modulo $N_{k}$) on the
tape of level $k$ macro-tiles, and
$\log k +O(\log N_{k})$ bits are required for that.
Both $\log k$ and $\log N_{k}$ should be
much less than $N_{k-1}$, so all the computations could fit in the
available time frame. This means that $N_k$ should not increase
too fast or too slowly. Say, $N_k=\log k$ is too slow (in this case $k$
occupies almost all available space in macro-tiles of level $k-1$, 
and we do not have enough time even for simple computations),
and
$N_{k}=2^{N_{k-1}}$ is too fast (in this case $\log N_{k}$ is too large compared
to  time and space available on the computation zone of a macro-tile
of level $k$). Also we need to compute
$N_{k}$ when $k$ is known, so we assume that not only the size
of $N_{k}$ (i.e., $\log N_{k}$) but also the time needed to
compute it (given $k$) is small compared to $N_{k-1}$. These
restrictions still allow many possibilities: Say, $N_k$ could be
proportional to
$\sqrt{k}$, $k$, $2^k$, $2^{(2^k)}$, or $k!$
  Note that we say ``proportional'' since $N_k$ needs to be
reasonably large even for small $k$ (we need some space in the macro-tile for wires and 
all our estimates for computation time are not precise but only asymptotic,
so we need some reserve for small $k$).

There is one more problem: It is not enough to ensure that the value of $k$
is the same for neighbor macro-tiles. We also need to ensure that this
value is correct, i.e., is $1$ for level $1$ macro-tiles made of $\tau_0$-tiles, is $2$
for level $2$ macro-tiles made of $\tau_1$-tiles, etc. To guarantee this,
we need to compare somehow the level information that is present in a
macro-tile and its sons.
Using the anthropomorphic
terminology, we say that each macro-tile ``knows'' its level, since
it is explicitly written on its tape, and this
is, so to say,  ``conscious''
information processed by a computation in the computation region
of the macro-tile. One may say also that a macro-tile of any level
contains ``subconscious'' information (``\emph{existing in the mind but
not immediately available to consciousness}''  \cite{dict}). This is the information that is conscious for  its sons, grandsons,
and so on (all the
way down to the ground level). The problem is that the macro-tile cannot
check consistency between conscious and subconscious information since
the latter is unavailable (the problem studied by psychoanalysis in a different
context).

The solution is to check consistency in the son, not in the
father. Every tile knows its level and
also knows its position in its father. So it knows whether it is in the place where its father
should keep level bits, and it can check whether indeed the level
bit that its father keeps in this place is consistent with the level
information the tile has. (In fact we used the same trick when we
simulated a substitution rule: A check that the father letter of
a tile coincides with the letter of the father tile is done in
the same way.)
The careful reader will also note here that now the neighbor tiles
will automatically have the same level information, so there is no
need to check consistency between neighbors.

This kind of  ``self-similar'' structure with variable zoom factors can be
useful in some cases. Though it is not  self-similar according
to our definition, one can still easily prove that any tiling is
aperiodic. Note that now the computation time for the Turing machine
simulated in the central part increases with level, and this can
be used for a simple proof of undecidability of the domino problem.
The problem in the standard proof (based on the self-similar
construction with fixed zoom factor) is that we need to place
computations of unbounded size into this self-similar structure,
and for that we need special geometric tricks
(see~\cite{durand-gurevich,berger}). With our new construction,
if we want to reduce an instance of the halting problem (some
machine $M$) to the domino problem, we add to the program
embedded in our construction the parallel computation of $M$ on
the empty tape; if it terminates, this destroys the tiling.

In a similar way we can show that the existence of a periodic tiling
is an undecidable property of a tile set, and, moreover,
the tile sets that admit periodic tilings and tile sets that have no
tilings form two inseparable sets (another classical result; see~\cite{gurevich-koryakov}).
Recall that two sets $A$ and $B$ are called
(computably) \emph{inseparable} if there is no computable set $C$
such that $A\subset C$ and $B\cap C=\emptyset$.

Here is an example of a more exotic version of the latter result
(which  probably is of no
interest in itself but just serves as an illustration of the technique). We
say that a tile set $\tau$ is \emph{$m$-periodic} if
$\tau$-tilings exist and for each of them the set of periods is
the set of \emph{all} multiples of $m$, in other words,
if the group
of periods is generated by $(0,m)$ and $(m,0)$. Let
$E$ [respectively $O$] be all $m$-periodic tile sets for all even $m$
[respectively odd $m$].

\begin{thm}
        \label{reduction}
The sets $E$ and $O$ are inseparable enumerable sets.
\end{thm}

\begin{proof} It is easy to see that the property ``to be an
$m$-periodic tile set'' is enumerable (both the existence
of an $m$-periodic tiling
and enforcing periods $(m,0)$ and $(0,m)$ are enumerable
properties).

It remains to reduce some standard pair of inseparable sets
(say, machines that terminate with output $0$ and $1$) to
$(E,O)$. It is easy to achieve this by using the technique explained above.
Assume that the numbers $N_k$ increase, being odd integers as
long as the computation of a given machine does not terminate.
When and if it terminates with output $0$ [respectively~$1$], we require
periodicity with odd [respectively~even] period at the next
level.\end{proof}

Another application of a variable zoom factor is the proof of
the following result obtained by Lafitte and Weiss
(see~\cite{lafitte-weiss}) using a Turing machine simulation
inside a Berger--Robinson construction.

\begin{thm}
Let $f$ be a total computable function whose arguments and values
are tile sets. Then, there exists a tile set $\tau$ that simulates
a tile set $f(\tau)$.
\end{thm}

Here we assume that some computable encoding for tile sets is
fixed. Since there are no restrictions on the computation
complexity of $f$, the choice of the encoding is not important.

\begin{proof} Note that for identity function $f$ this result
provides the self-simulating tile set of
Section~\ref{simulating-itself}. To prove it in the general case, we may use the same
kind of  fixed-point technique. However, there is a problem:
The computation resources inside a tile are limited (by its
size) while time needed to compute $f$ can be large (and,
moreover, depends on the tile size).

The solution is to postpone the simulation to large levels. If a
tile set $\tau_0$ simulates $\tau_1$, which simulates $\tau_2$,
which simulates, etc., up to $\tau_n$, then $\tau_0$ simulates
$\tau_n$, too. Therefore we may proceed as follows.

We use the construction explained above with a variable zoom
factor. Additionally, at each level the computation starts with
a preliminary step that may occupy up to (say) half of the
available time. On this step we read the program
that is on the tape and convert it into the tile set. (Recall that
each program determines some tile set $\tau_0$ such that
$\tau_0$-tilings can be uniquely split into macro-tiles, and
this program is written on a read-only part of the tape
simulated in the computation zone of all macro-tiles, as  was explained
in Section~\ref{simulating}.)  Then, we
apply $f$ to the obtained tile set.

This part of the computation checks also that it does not use
more than half of the available time and that the output is
small enough compared to the macro-tile size. If this time turns out
to be insufficient or the output is too big, this part is
dropped and we start a normal computation for the variable zoom
factor, as explained above. In this case, the zoom factor on the next level should be
greater than the zoom factor on the current level (e.g., we may assume
$N_k=Ck$ for some large enough constant $C$).
However, if the time is large enough and
the result (the list of tiles that corresponds to $f$'s output) is small
compared to the macro-tile size, we check that the macro-tile (of the
current level) belongs to the tile set computed.
The hierarchy of macro-tiles stops at this level. The behavior of macro-tiles
at this level depends on $f$: They are isomorphic to $f(\tau_0)$-tiles.
Since the program is the same at all levels and the computation
of $f$ should be finite (though may be very long), at some (big
enough) level the second possibility is activated, and we get
a macro-tile set isomorphic to $f(\tau)$, where $\tau$ is the tile set
on the ground level.\end{proof}

\bigskip

Another application of the variable zoom factor technique is the construction of tile sets with any
given computable density. Assume that a tile set is given and,
moreover, that all tiles are divided into two classes, say, $A$-tiles
and $B$-tiles. We are interested in a fraction of $A$-tiles in a
tiling of an entire plane or its large region. If the tile set
is flexible enough, this fraction can vary. However, for some
tile sets this ratio tends to a limit value when the size of a
tiled region increases. This phenomenon is captured in the
following definition: We say that tile set $\tau$ divided into
$A$- and $B$-tiles \emph{has a limit density $\alpha$} if for every
$\varepsilon>0$ there exists $N$ such that for any $n>N$ the fraction of $A$-tiles in any tiling of the $n\times n$ square is
between $\alpha-\varepsilon$ and $\alpha+\varepsilon$.

\begin{thm}

\textup{\textbf{(i)}} If a tile set has a density $\alpha$, then $\alpha$ is a
computable real number in $[0,1]$.
\textup{\textbf{(ii)}}~Any computable real number $\alpha\in[0,1]$ is a
density of some tile set.

\end{thm}

\begin{proof} The first part of the proof is a direct corollary of the definitions.
For each $n$ we
can consider all tilings of the $n\times n$ square and look for the
minimal and maximal fractions of $A$-tiles in them. Let us denote the minimal and maximal fractions
by $m_n$ and $M_n$ respectively. These rational numbers are computable given $n$.
It is easy to see that the limit frequency
(if it exists) is in the interval $[m_n, M_n]$. Indeed, in a large square
split into squares of size $n\times n$ the fraction of $A$-tiles is between
$m_n$ and $M_n$, being at the same time arbitrarily close to $\alpha$.
Therefore, $\alpha$ is computable (to get its value with precision $\varepsilon$, we increase $n$ until the difference
between $M_n$ and $m_n$ becomes smaller than $\varepsilon$).

It remains to prove~(ii).  Since $\alpha$ is computable,
there exist two computable sequences of rational numbers $l_i$
and $r_i$ that converge to $\alpha$ in such a way that
        $$
[l_1,r_1]\supset[l_2,r_2]\supset[l_3,r_3]\supset\cdots.
        $$
Our goal will be achieved if macro-tiles of the first level
have density of either $l_1$ or $r_1$, macro-macro-tiles
of the second level
have density of either $l_2$ or $r_2$, and so on. Indeed,
each large square can be split into macro-tiles (and the
border that does not change the density much), so in any large
square the fraction of $A$-tiles is (almost) in $[l_1,r_1]$. The same
argument works for macro-macro-tiles, etc.

However, this plan cannot be implemented directly. The main difficulty is that the computation of $l_i$ and $r_i$
may require a lot of time whereas the computation abilities of
macro-tiles of level $i$ are limited. (We use variable zoom
factors, e.g., we may let $N_k=Ck$,
 but they cannot grow too fast.)

The solution is to postpone the switch from densities $l_i$ and
$r_i$ to densities $l_{i+1}$ and $r_{i+1}$ to the higher
level of the hierarchy where the computation has enough
time to compute all these four rational numbers and
find out in which proportion $l_i$- and $r_i$-tiles should be
mixed in $l_{i+1}$- and $r_{i+1}$-tiles. (We need the denominators in both fractions 
$l_{i+1}$ and $r_{i+1}$ to be equal to the number
of $i$-level macro-tiles in the $(i+1)$-level macro-tile,
but this restriction can  always be satisfied by a slight change
in the sequences $l_k$ and $r_k$, which leaves $\alpha$ unchanged.)
So, we allocate, say, the first half of the available time for
a controlled computation of all these values; if the computation
does not finish in time, the densities for the next level are
the same as for the current level. 
(We require that all macro-tiles in the same father tile have the same density, either $l_i$ or $r_i$).
If the computation terminates
in time, we use the result of the computation to have two
types of the next level tiles: one with density $l_{i+1}$ and
one with density $r_{i+1}$. They are made by using prescribed
amounts of $l_i$- and $r_i$-tiles. (Since each tile knows its
coordinates, it can find out whether it should be of the first
or second type.) This finishes the construction.\end{proof}

\section{Strongly aperiodic tile sets revisited}
\label{strongly}

In Section~\ref{thue} we constructed a tile set such that every
tiling is $\alpha$-aperiodic for every $\alpha<1/4$. Now we
want to improve this result and construct a tile set such that
every tiling is, say, $0.99$-aperiodic (here $0.99$ can be
replaced by any constant less than~$1$). It is easy to see that
this cannot be achieved by the same argument, with Thue--Morse
substitutions, nor with any substitutions in a two-letter
alphabet; we need a large alphabet to make the constant close to
$1$.

It is possible to achieve $0.99$-aperiodicity with a carefully chosen
substitution rule (in a bigger alphabet), as recently proposed
by Pritykin and Ulyashkina~\cite{pri-ul},
by just applying Theorem~\ref{substitution-tiling} (similarly to the argument
for the Thue--Morse substitution presented in Section~\ref{thue}).
In this section we present an alternative proof of this result.
We  exploit
substitution rules with variable zoom factors (and different
substitutions on each level) and use the idea of an error-correcting code.

Instead of one single alphabet, $A$, we now consider an infinite sequence of
finite alphabets, $A_0, A_1, A_2,\ldots$; the cardinality of
$A_k$ will grow as $k$ grows. Then, we consider a sequence of
mappings:
        $$
s_1\colon A_1 \to A_0^{N_0\times N_0},\quad
s_2\colon A_2 \to A_1^{N_1\times N_1},\quad
s_3\colon A_3 \to A_2^{N_2\times N_2},\ldots,
        $$
where $N_0, N_1, N_2,\ldots$ are some positive integers
(zoom factors); $N_k$ will  increase as $k$ increases.

Then, we can compose these mappings. For example, a letter $z$ in
$A_2$ can be first replaced by an $N_1\times N_1$ square $s_2(z)$
filled by $A_1$-letters. Then, each of these letters can be
replaced by an $N_0\times N_0$ square filled by $A_0$-letters
according to $s_1$, and we get an $N_0N_1\times N_0N_1$ square
filled by $A_0$-letters; we denote this square by $s_1(s_2(z))$
(slightly abusing the notation).

We call all this (i.e., the sequence of $A_k$, $N_k$, $s_k$) a
\emph{substitution family}. Such a family defines a class of
$A_0$-configurations compatible with it (in the same way as in
Section~\ref{substitution-tiling}). Our plan is to construct a
substitution family such that

\begin{itemize}
\item every configuration compatible with this family is
$0.99$-aperiodic, and

\item there exists a tile set and projection of it onto $A_0$
such that only compatible configurations (and all compatible
configurations) are projections of tilings.
\end{itemize}

In other words, we use the same argument as before (proving
Theorem~\ref{aperiodic}) but use a substitution family instead
of one substitution rule. This substitution family will have two
special properties:

\begin{description}

\item{A.} Symbols used in different locations are different.
This means that $A_k$-letters that appear in a given position of
the squares $s_{k+1}(z)$ for some $z\in A_{k+1}$ never appear in any
other places of these squares (for any $z$); thus, set $A_k$ is split into
$N_k\times N_k$ disjoint subsets used for different positions in
$N_k\times N_k$ squares.

\item{B.} Different letters are mapped to squares that are far
away in terms of Hamming distance. This means that if $z,w\in A_{k+1}$
are different, then the Hamming distance between images $s_{k+1}(z)$ and $s_{k+1}(w)$
is large: The fraction of positions in
the $N_k\times N_k$ square, where $s_{i+1}(z)$ and $s_{i+1}(w)$
have equal letters does not exceed $\varepsilon_k$.

\end{description}

Here $\varepsilon_i$ will be a sequence of positive reals such that
$\sum_{i\ge0} \varepsilon_i <0.01$.

\medskip

This implies that composite images of different letters are also
far apart. For example, the fraction of positions in the $N_0
N_1\times N_0N_1$ square where $s_1(s_2(z))$ and $s_1(s_2(w))$
coincide does not exceed $\varepsilon_0+\varepsilon_1<0.01$.
Indeed, in $s_2(z)$ and $s_2(w)$ we have at most
$\varepsilon_1$-fraction of matching letters; these letters
generate $\varepsilon_1$-fraction of matching $A_0$-letters on
the ground level; all other (nonmatching) pairs add
$\varepsilon_0$-fraction. In fact, we get even a stronger bound
$1-(1-\varepsilon_0)(1-\varepsilon_1)$.

For the same reasons, if we take two different letters in $A_k$ and
then drop to the ground level and obtain two squares of size
$N_0N_1\cdots N_{k-1}\times N_0N_1\cdots N_{k-1}$ filled by
$A_0$-letters, the fraction of coincidences is at most
$\varepsilon_0+\cdots+\varepsilon_{k-1}<0.01$.

This property of the substitution family implies the desired
property:

\begin{lem}
If an $A_0$-configuration $U$ is compatible with  a substitution
family having properties \textup{(A)} and \textup{(B)}, then $U$ is $0.99$-ape\-ri\-o\-dic.
\end{lem}

\begin{proof} Consider a shift vector $T$. If $T$ is not a
multiple of $N_0$ (one of the coordinates is not a multiple of
$N_0$), then property (A) guarantees that the original configuration
and its $T$ shift differ everywhere. Now assume that $T$ is a
multiple of $N_0$. Then, $T$ induces a $(T/N_0)$-shift of an
$A_1$-configuration $U_1$ that is an $s_1$-preimage of $U$. If
$T$ is not a multiple of $N_0N_1$, then $T/N_0$ is not a
multiple of $N_1$ and for the same reason this $(T/N_0)$-shift
changes all the letters in $U_1$. Different letters in $A_1$
are mapped to $N_0\times N_0$ squares that coincide in at most
$\varepsilon_0$-fraction of positions.

If $T$ is a multiple of $N_0N_1$ but not $N_0N_1N_2$, we get a
$T/(N_0N_1)$ shift of $A_2$-con\-fi\-gu\-ra\-tion $U_2$ that changes all
its letters, and different letters give squares that are
$1-(\varepsilon_0+\varepsilon_1)$ apart. The same argument works
for the higher levels.\end{proof}

It remains to construct a substitution family that has
properties (A) and (B) and can be enforced by a tile set. Property (B) (large Hamming distance) is standard for coding
theory, and the classical tool is the  Reed--Solomon code.

Let us recall the idea of the Reed--Solomon code (for details see,
e.g., \cite{code-book}).
The codewords of the Reed--Solomon code are
tables of (values of) polynomials of bounded degree.
More precisely, we fix some finite field $\mathbb{F}_q$ of size $q$
and an integer $d>0$. Let $p(x)=a_0+a_1x+\cdots+a_{d-1}x^{d-1}$
be a polynomial over $\mathbb{F}_q$ of degree less than $d$. Then
the codeword corresponding to $p(x)$ (i.e., the encoding of
the sequence $a_0,\ldots,a_{d-1}$) is a vector in $(\mathbb{F}_q)^q$
(i.e., a sequence of $q$ elements of the field),
which consists of the values of this polynomial
computed at all points $x\in \mathbb{F}_q$.
Thus, for given parameters $d$ and $q$, the code consists of $q^d$ codewords.
Since two polynomials of degree less than $d$
can coincide in at most $(d-1)$ points, the distance between any two
codewords is at least $q-d+1$. Of course, this construction can
be used even if the desired length of the codewords is not a size of any
finite field; we can choose a slightly larger field and use only part
of its elements.

Now we embed these codes in a family of substitution rules.
First, let $B_k$ be a finite field (the size of which is specified below)
and let $A_k$ be equal to $B_k\times
\{0,1,\ldots,N_k-1\}\times\{0,1,\ldots,N_k-1\}$; let us agree that
we use letters $\langle b,i,j\rangle$ only in the $(i,j)$-position
of an $s_{k+1}$-image. This  trivially implies requirement (A).

Then, we construct a code that encodes each $A_{k+1}$-letter $w$
by  a string of length $N_k^2$ made of $B_k$-letters (arranged
in a square); adding the coordinates, we get the $s_{k+1}$-image of
$w$. Thus, we need a sequence of codes:
        \begin{align*}
s_1 : A_1 \to B_0^{N_0\times N_0},   \quad 
         & \text{such that } s_1(w) \text{ and}\ s_1(w')\  \text{coincide at most in $\varepsilon_0$ fraction}\\
         & \text{of all positions (if $w\ne w'$),}\\
s_2 : A_2 \to B_1^{N_1\times N_1},
        \quad & \text{such that }
         s_2(w) \text{ and } s_2(w')\ \text{coincide at most in $\varepsilon_1$ fraction}\\
         & \text{of all positions (if $w\ne w'$),}\\
        &\ldots
.        \end{align*}
To satisfy requirement (B), we need  a code with the Hamming distance (between every two codewords)
at least
$(1-\varepsilon_{k})N_{k}^2$. The Reed--Solomon code works well here. The size of the
field can be equal to the length of the codeword, i.e.,
$N_k^2$. Let us decide that $N_k$ is a power of $2$ and the size
of the field $B_k$ is exactly $N_k^2$.
(There are fields of size $2^t$ for every $t=1,2,3,\ldots$; we could also use
$\mathbb{Z}/p\mathbb{Z}$ for prime $p$ of an appropriate size.)
To achieve the required code distance, we use polynomials
of degree less than $\varepsilon_{k}N_k^2$. The number of
codewords (polynomials of degree less than $\varepsilon_{k}N_k^2$) is at least
$2^{\varepsilon_k N_k^2}$ (even if we use only polynomials with coefficients $0$ and $1$). This  is enough if
        $$
|A_{k+1}|\le
2^{\varepsilon_k N_k^2}.
        $$
Recalling that $|A_{k+1}|=|B_{k+1}|\cdot N_{k+1}^2$ and that
$B_{k+1}$ is a field of size $N_{k+1}^2$, we get the
inequality
        $$
    N_{k+1}^4 \le 2^{\varepsilon_k N_k^2}, \text{\ \ or \ \ }
   4\log N_{k+1} \le \varepsilon_k N_k^2.
        $$
Now let $N_k=2^{k+c}$ for some constant $c$; we see that for
large enough $c$ this inequality is satisfied for
$\varepsilon_k$ with sum less than $0.01$ (or any other
constant), since the left-hand side is linear in $k$ while the
right-hand side is exponential.

Now it remains to implement all this scheme using tiling rules. As we
have discussed, the zoom factor $N_k=2^{k+c}$ is acceptable for the
construction. This factor leaves enough space to keep on the tape two
substitution letters (for the tile itself and its father tile),
since these letters require linear size (in $k$). Moreover, we
have enough time to perform the computations
in the finite fields needed to construct the error-correction
code mappings.  Indeed, in a $k$-level macro-tile we are allowed to use
exponential (in the bit size of the field element) time. Recall that one can operate with elements in the field of size $2^r$ using
polynomial (in $r$) time; to this end, we need to construct some irreducible
polynomial $p$ of degree $r$ over the field of two elements and then perform
arithmetic operations (on polynomials) modulo $p$. All these operations can be
done by deterministic algorithms in polynomial time  (see, e.g., \cite{finite-fields}).
Thus, we can reuse here the construction of the proof of Theorem~\ref{substitution-tiling}.

The construction above works with every constant $\alpha<1$ instead of $0.99$.
So, we get a stronger version of Theorem~\ref{aperiodic}:
\begin{thm}
        \label{aperiodic-strong}
For every $\alpha<1$
there exists a tile set $\tau$ such that $\tau$-tilings exist
and every $\tau$-tiling is $\alpha$-aperiodic.
\end{thm}

\textbf{Remark}. We can also get an $\alpha$-aperiodic
 tile set  (for every $\alpha<1$)
as a corollary of the result of the next section; indeed, we construct
there a tile set such that any tiling embeds a horizontal
sequence with high-complexity substrings, and such a sequence
cannot match itself well after a shift (in fact, to get $\alpha$-aperiodicity
we would need to
replace a binary alphabet by a larger finite alphabet in this
argument). We can superimpose this with a similar
$90^\circ$-rotated construction; then, any nonzero translation will
shift either a vertical or a horizontal sequence and therefore
change most of the positions. Note that in this way we can also
get a tile set that is $\alpha$-far from every periodic pattern (a
slightly different approach to defining ``strong aperiodicity'').
However, the arguments used in Section~\ref{complex} are more complicated
than the proof of this section. So we preferred to present here
a simpler and more direct proof of Theorem~\ref{aperiodic-strong}.

\section{Tile sets with only complex tilings}
\label{complex}

In this section we provide a new proof of the following result
from~\cite{dls}:

\begin{thm}
\label{thm:complex}
There exist a tile set $\tau$ and constants $c_1>0$ and $c_2$
such that $\tau$-tilings exist and in every $\tau$-tiling $T$
every $N\times N$ square has Kolmogorov complexity at least
$c_1N-c_2$.
\end{thm}

Here Kolmogorov complexity of a tiled square is the length of the
shortest program that describes this square. We assume that
programs are bit strings. Formally speaking, Kolmogorov complexity
of an object depends on the choice of programming language.
(Consult~\cite{uppsala-notes} for the definition and properties of Kolmogorov
complexity.)
However, in our case the choice of programming language
does not matter, and you may think of Kolmogorov complexity
of an object as the length of the shortest program in your favorite programming
language that prints out this object.
We need to keep in mind only two important
properties of Kolmogorov complexity.
First, the Kolmogorov complexity
function is not computable, but it is \emph{upper semicomputable}. This
means that there is an algorithm that for a given $n$ enumerates
all objects that have complexity less than $n$.
The enumeration can be done by a brute force search over all short
descriptions. We cannot say in advance which programs stop with some output and which do not,
but we can run all programs of length less than $n$ in parallel, and enumerate the list of their outputs,
as some programs terminate.
Second, any computable transformation (e.g., the change of encoding)
changes Kolmogorov complexity at most by $O(1)$.
We refer to~\cite{dls} for a discussion of Theorem~\ref{thm:complex} (why it
is optimal, why the exact value of $c_1$ does not matter, etc.)
and other related results.

\subsection{A biinfinite bit sequence}

\begin{proof}
We start the proof in the same way as in~\cite{dls}: We assume
that each tile keeps a bit that propagates (unchanged) in the
vertical direction. Then, any tiling contains a biinfinite
sequence of bits $\omega_i$ (where $i\in \mathbb{Z}$). Any
$N\times N$ square contains an $N$-bit substring of this string,
so if (for large enough $N$) every $N$-bit substring of $\omega$
has complexity at least $c_1 N$ for some fixed $c_1$, we are done.

We say that a sequence $\omega$ has \emph{Levin's property}
if every $N$-bit substring $x$ of $\omega$ has complexity
$\Omega(N)$. Such a biinfinite sequence indeed exists (see~\cite{dls};
another proof can be obtained by using the Lovasz local lemma;
see~\cite{rumush}). So our goal is to formulate tiling rules in
such a way that a correct tiling ``ensures'' that the biinfinite
sequence embedded in it indeed has this property.

The set of all ``forbidden'' binary strings, i.e., strings $x$ such that $K(x)<c_1|x|-c_2$ (where $K(x)$ denotes the Kolmogorov
complexity of $x$, and $|x|$ denotes the length of~$x$) is
enumerable: There is an algorithm that generates the list of all forbidden
substrings. It would be nice to embed into the tiling a
computation that runs this algorithm and compares its output
strings with the substrings of $\omega$; such a computation
blows up (creates a tiling error) if a forbidden substring is
found.

However, there are several difficulties.

\begin{itemize}

\item

Our self-similar tiling contains only finite
computations. The higher is rank $k$, the bigger are the $k$-level macro-tiles, 
and the longer computations they can contain.
But at any level the computation remain finite.
This is a problem since for a given string $x$ we do not
know \emph{a priori} how much time the shortest program for $x$ uses,
so we never can be sure that the Kolmogorov
complexity of $x$ is large. Hence, each substring of $\omega$ should be
examined in computations somehow distributed over infinitely many macro-tiles.

\item

The computation at some level deals with bits encoded in the
cells of that level, i.e., written on the computation tape. So the
computation cannot access the bits of the sequence (that are
``deep in the subconscious'') directly and some mechanism to dig
them out is needed.

\end{itemize}

Let us explain how to overcome these difficulties.

\subsection{Delegation of bits}\label{bit-delegation}

A macro-tile of level $k$ is a square whose side is $L_k=N_0\cdot
N_1 \cdots N_{k-1}$, so there are $L_k$ bits of the
sequence that intersect this macro-tile. Let us delegate each of
these bits to one of the macro-tiles 
of level $k$ it intersects. (We do it for every $k$.)  Note that
the macro-tile of the next level is made of $N_k\times N_k$
macro-tiles of level $k$. We assume that $N_k$ is much bigger than $L_k$ (see the end of this subsection for more details on the choice of $N_k$); this guarantees
that there are enough macro-tiles of level $k$ (in the next level
macro-tile) to serve all bits that intersect them. Let us decide
that the $i$th (from bottom to top) macro-tile of level $k$  in a
$(k+1)$-level macro-tile serves (consciously knows, so to say) the
$i$th bit (from the left) in its zone (see Fig.~\ref{fpt.8.mps}). Since $N_k\gg L_{k}$,
we have many more macro-tiles of level $k$  (inside some macro-tile
of level $k+1$) than needed to serve all bits. So some $k$-level macro-tiles
remain unused.

\begin{figure}[h]
 \begin{center}
        $$
\includegraphics[scale=1.0]{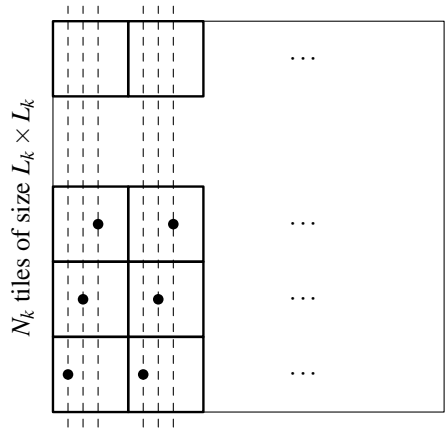}
        $$
 \end{center}
 \caption{Bit delegation.}
\label{fpt.8.mps}
\end{figure}

Thus, each bit (each vertical line) has a representative on every
level---a macro-tile that consciously knows this bit. However,
we need some mechanisms that guarantee that this information is
indeed true (i.e., consistent on different levels). On the bottom
level this is easy to achieve, since the bits are available directly.

To guarantee the consistency we use the same trick as in
Section~\ref{substitution}: At each level a macro-tile keeps
not only its own bit but also its father's bit, and makes
necessary consistency checks. Namely, each macro-tile knows
(has on its computation tape):

\begin{itemize}

\item the bit delegated to this macro-tile;

\item the coordinates of this macro-tile in its father
        macro-tile (which are already used in the fixed-point
        construction); note that the $y$-coordinate is at the same time the
        position of the bit delegated to this macro-tile
        (relative to the left boundary of the macro-tile);

\item the bit delegated to the father of this macro-tile;
and
\item the coordinates of the father macro-tile in the
        grandfather macro-tile.

\end{itemize}

This information is subject to consistency checks:

\begin{itemize}

\item The information about the father macro-tile should coincide
       with the same information in the neighbor tiles (unless they
       have a different father, i.e., one of the coordinates is
       zero).

\item If  the bit delegated to the father
       macro-tile 
       is from the same vertical column as the bit delegated for this macro-tile, these two bits
       should match.

\item If the macro-tile occupies a place in its
      father macro-tile where the bit delegated to
      the father or some bits of the father's coordinates
      (inside the grandfather macro-tile)   are kept, then this partial information on the
      father level should be should be consistent with the information
      about father coordinates and bit.

\end{itemize}

These tests guarantee that the information about the father is the
same in all brothers, and some of these brothers (which are
located on the father tape) can check it against actual father
information; at the same time some other brother (that has the
same delegated bit as the father) checks the consistency of the
delegated bits information.

Note that this scheme requires that not only $\log N_k$ but also
$\log N_{k+1}$ is much less than $N_{k-1}$. This requirement,
together with the inequality $L_k=N_0 N_1\cdots N_{k-1}\le  N_k$ (discussed earlier),
is satisfied if $N_k=Q^{c^k}$, where $Q$ is a large enough constant (which is needed also to
make macro-tiles of the first level large enough)
and $c>2$ (so $1+c+c^2+\cdots+c^{k-1}<c^k$).

Later, in Section~\ref{robust-complex}, the choice of $c$ has to be reconsidered: We need $2<c<3$
to achieve error correction, but for our current purposes this does not matter.

\subsection{Checking bit blocks}

We explained how macro-tile of any level can have  true information about one bit (delegated to it). However, we need to
check not bits but substrings (and artificially introduce a tiling error if a 
forbidden string appears). Note that it is acceptable to test only very
short substrings compared to the macro-tile size ($N_k$). If
this test is done on all levels, this  restriction
does not prevent us from detecting any violation. (Recall that short
forbidden substrings can appear very late in the generation
process, so we need computation at arbitrary high levels for
this reason, too.)

So we need to provide more information to macro-tiles. This can be done
in the following way. Let us require that a macro-tile contains not one
bit but \emph{a group of bits to check}:
 a group of bits that starts at the delegated bit and has
length depending on the level $k$ (and growing very slowly with
$k$; e.g., $\log\log\log k$ is slow enough). 
If this group is not completely inside a macro-tile
(i.e., it extends out of the region occupied by the
macro-tile), we ignore the outstanding part.

Similarly, a macro-tile should have this information for the
father macro-tile (even if the bits are outside its own region).
This information about the father macro-tile should be the same for brothers
(which is checked by matching macro-colors of neighboring brothers).
Also each macro-tile checks (on its computational zone) that the value of
its own \emph{delegated bit}  is coherent with
its father's string of \emph{bits to check}:  A macro-tile knows its coordinate
in the father macro-tile and the coordinates of the father tile in the grandfather,
so it knows whether its delegated bit makes a part of the father's bits to check.

The computation in the computation zone generates the list of all forbidden strings
(strings that have too small  Kolmogorov complexity)
and checks the generated forbidden strings 
against all the substrings of the group of bits available
to this macro-tile. This process is bounded in time and space,
but this does not matter since every string is considered on a
high enough level.

Our construction has a kind of duplication: We first guarantee
the consistency of information for individual bits, and then, we do the same
for substrings. The first part of the construction is still needed,
since we need arbitrarily long substrings to be checked by macro-tiles (of high enough level); thus
 delegation of substrings cannot start from the ground level where the tile size
is limited, so we need to deal with bits separately.

\subsection{Last correction}\label{last-correction}

The argument just explained  still needs some correction. We
claim that every forbidden string will be detected at some level
where it is short enough compared to the level parameters.
However, some strings may never become a part of one
macro-tile. Imagine that there is some vertical line that is a
boundary between macro-tiles of all levels (so we have bigger
and bigger tiles on both sides, and this line is still the
boundary between them; see Fig.~\ref{degenerate.mps}). Then, a substring that crosses this line
will never be  checked and therefore we cannot guarantee that it
is not forbidden.
\begin{figure}
\begin{center}
\includegraphics[scale=0.6]{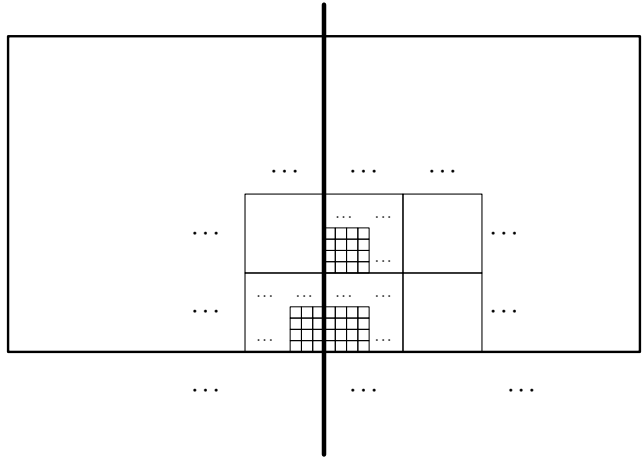}
\end{center}
\caption{Degenerate case: An infinite vertical line  is a boundary between macro-tiles of all levels.}
\label{degenerate.mps}
\end{figure}

There are several ways to get around this problem. One can
decide that each macro-tile contains information not only about
blocks inside its father macro-tile but in a wider region (say,
three times wider, including ``uncle'' macro-tiles); this information
should be checked for consistency between ``cousins'', too.
\label{discussion-degenerated-case}
This trick (extending  zones of responsibility
for macro-tiles) will be used later in Section~\ref{subshifts}.

But to prove Theorem~\ref{thm:complex}
a simpler solution is enough. Note that even if a string on
the boundary is never checked, its parts (on both sides of the
boundary) are, so their complexity is proportional to their
length. One of the parts has length at least half of the
original length, so we still have a complexity bound, though the
constant will be twice smaller.

This finishes the proof of Theorem~\ref{thm:complex}.\end{proof}

\section{Subshifts}
\label{subshifts}

The analysis of the proof in the previous section shows that it can be divided into two parts. We defined \emph{forbidden} strings as bit strings that are sufficiently long and have complexity at most ${\alpha\cdot(\text{length}})$. We started by showing that biinfinite strings without forbidden factors (substrings) exist. Then, we constructed a tile set that embeds such a biinfinite string in every tiling.

The second part can be separated from the first one, and in this way we get new proofs for some results of Simpson~\cite{simpson-subshifts}  and Hochman~\cite{hochman} about effectively closed subshifts.

Fix some alphabet $A$. Let $F$ be a set of $A$-strings. Consider a set $S_F$ of all biinfinite  $A$-sequences that have no factors (substrings) in $F$. This set is a \emph{closed one-dimensional subshift} over $A$, i.e., a closed shift-invariant subset of the space of all biinfinite $A$-sequences.  If the set $F$ is (computably) enumerable, $S_F$ is called an \emph{effectively closed one-dimensional subshift} over $A$. If $F$ is finite, $S_F$ is called a \emph{subshift of finite type}.

We can define two-dimensional subshifts in a similar way.
More precisely, let $F$ be a set of two-dimensional patterns (squares filled with $A$-letters). Then, we can consider a set $S_F$ of all $A$-configurations (= mappings $\mathbb{Z}^2\to A$) that do not contain any pattern from $F$. This is a closed shift-invariant set of $A$-configurations (= two-dimensional closed subshift over $A$). If $F$ is (computably) enumerable, $S_F$ is called \emph{a two-dimensional effectively closed subshift} over $A$. If $F$ is finite, $S_F$ is called \emph{a two-dimensional subshift of finite type}.

As we have mentioned, subshifts of finite type (interpreted as local rules) are closely related to tilings. Each tile set determines a subshift where $A$ is the set of tiles and forbidden patterns are pairs of neighbor nonmatching tiles. Going in the other direction, we should be more careful. A tile set in our definition cannot contain two different tiles with exactly the same colors. This leads to some problems. For example, the full shift over a two-letter alphabet (i.e., the set of all biinfinite sequences over a two-letter alphabet)  cannot be represented by a set of two tiles. However, any subshift of  finite type can be represented by some tile set. More precisely,
for any subshift $S$ of finite type over alphabet $A$  there is a tile set $\tau$ and some mapping $E\colon\tau\to A$ such that $E$ induces a \emph{bijection} between the set of all $\tau$-tilings and the set of all configurations of the subshift $S$: we apply $E$ pointwise to a $\tau$-tiling and get some $A$-configuration from the subshift; for each configuration in the subshift there exists exactly one $\tau$-tiling in the $E$-preimage of this configuration. 
 Such a tile set can be constructed as follows: Tiles are squares of large enough size filled by $A$-letters (a square with no patterns forbidden for this subshift); each tile represents a part of the configuration, and side colors are used to ensure that neighbor tiles overlap correctly. The mapping extracts (say) the central letter from a square.

Thus, subshifts of finite type and tilings are essentially the same kind of objects. On the other hand, the effectively closed subshifts of dimension make  a more general class of objects than subshifts of finite type. E.g., every nonempty one-dimensional subshift of finite type must contain a periodic configuration; for  one-dimensional effectively closed subshifts this is not the case. However, the following theorem shows that two-dimensional subshifts of finite type are powerful enough to simulate any effectively closed one-dimensional subshift in the following sense (i.e., to simulate an effectively closed subshift, we need a subshift of finite type of dimension higher by $1$):

\begin{thm}
Let $A$ be some alphabet and let $S$ be a one-dimensional effectively closed subshift over $A$. Then, there exist an alphabet $B$, a mapping $r\colon B\to A$, and a two-dimensional subshift $S'$ of finite type over $B$ such that $r$-images of configurations in $S'$ are \textup(exactly\textup) elements of $S$ extended vertically \textup(vertically aligned cells contain the same $A$-letter\textup).
\end{thm}

(As we have mentioned, this result was independently obtained by Aubrun and Sablik using Robinson-style aperiodic tilings~\cite{sablik-aubrun}.)

\begin{proof}
The proof uses the same argument as in Theorem~\ref{thm:complex}. Each cell now contains an $A$-letter that propagates vertically.  Computation zones in macro-tiles generate (in available space and time) elements of the enumerable set of forbidden $A$-substrings and compare them with $A$-substrings that are made available to them. It remains to note that tiling requirements (matching colors) are local; that is, they define a finite type two-dimensional subshift.

Note that now the remark of Section~\ref{last-correction} (the trick of extension of zones of responsibility for macro-tiles) becomes crucial, since otherwise the image of a configuration from $S'$ may be a concatenation of two sequences (a left-infinite one and a right-infinite one); neither sequence  contains forbidden patterns but forbidden patterns may appear at the point of concatenation.\end{proof}

A similar argument shows that every two-dimensional effectively closed subshift can be represented as an image of a three-dimensional subshift of finite type (after a natural extension along the third dimension), any three-dimensional effectively closed subshift is an image of a four-dimensional subshift of finite type, etc.

This result is an improvement of a similar one proved by Hoch\-man (Theorem 1.4 in~\cite{hochman}, where the dimension increases by $2$), thus providing a solution of Problem~9.1 from~\cite{hochman}. Note also that it implies the result of~Simpson~\cite{simpson-subshifts} where one-dimensional sequences are embedded into two-dimensional tilings but in some weaker sense (defined in terms of Medvedev degrees).

One can ask whether a dimension reduction is essential here. For example, is it true that every two-dimensional effectively closed subshift is an image of some two-dimensional subshift of finite type?
The answer to this question (as well as related questions in higher dimensions) is negative. This follows from an upper bound in~\cite{dls} saying that every tile set (unless it has no tilings at all) has a tiling such that all $n\times n$ squares in it have complexity $O(n)$ (a result that immediately translates for subshifts of finite type) and a result from~\cite{rumush} that shows that some nonempty effectively closed two-dimensional subshift has $n\times n$ squares of complexity $\Omega(n^2)$. Therefore the latter cannot be an image of the first one (complexity can only decrease when we apply an alphabet mapping).

\section{Random errors}
\subsection{Motivation and discussion}
\label{random}

In what follows we discuss tilings with faults.  This means that
there are some places (faults) where colors of neighbor tiles do not match. We are interested in
``robust'' tile sets: those that maintain some structure (for example, can be
converted into an error-free tiling by changing a small fraction of tiles) if faults are sparse.

There are two almost equivalent ways to define faulty tilings.
We can speak about \emph{errors}
(places where two neighbor tiles do not match) or \emph{holes} (places
without tiles).  Indeed, we can convert a tiling error into a
hole (by deleting one of two nonmatching tiles) or convert
a one-tile hole (one missing tile)
into a small number of errors  (at most $4$) by placing an
arbitrary tile there. Holes look more natural if we start
with a set of holes and then try to tile the rest; however, if we imagine some process similar to crystallization when
a tiling tries to become correct by some trial-and-error
procedure, it is more natural to consider tiling errors. Since
it  makes little difference from the mathematical point
of view, we use both metaphors.

We use a hierarchical approach to hole patching that can be traced back
to~G\'acs, who used it in a much more complicated
situation~\cite{gacs}. This means that first we try to patch
small holes that are not too close to each other (by changing
small neighborhoods around them). This (if we are lucky enough)
makes larger (and still unpatched) holes more isolated since
there are fewer small holes around. Some of these larger holes
(which are not too large and not too close to each other) can be
patched again. Then, the same procedure can be repeated again for
the next level. Of course, we need some conditions (that guarantee
that holes are not too dense)
to make this procedure successful. These conditions are
described later in full detail, but the important question is the following:
How do we ensure that these conditions are reasonable (i.e., general
enough)? Our answer is as follows: We prove that if holes are generated at
random (with each position becoming a hole independently of other
positions with small enough probability $\varepsilon$), then the
generated set satisfies these conditions with probability~$1$.

From the physics viewpoint, this argument sounds rather weak. If
we imagine some crystallization process, errors in different
positions are not independent at all. However, this approach
could be a first approximation until a more adequate one is
found.

\medskip

Note that patching holes in a tiling could be considered as a
generalization of  percolation theory. Indeed, let us
consider a simple tile set made of two tiles: one with all black
sides and the other with all white sides. Then, the tiling conditions
reduce to the following simple condition: Each connected
component of the complement to the  set of holes is either
completely black or completely white. We want to make small
corrections in the tiling that patch the holes (and therefore
make the entire plane black or white). This means that initially
either we have small black ``islands'' in a white ocean or vice
versa, which is exactly what percolation theory says (it
guarantees that if holes are generated at random independently
with small probability, the rest consists of one large connected
component and many small islands.)

This example shows also that simple conditions such as low
density (in the Besicovitch sense) of the hole set are not enough.
A regular grid of thin lines can have low density but still
splits the plane into nonconnected squares; if half of these
squares are black and the others are white, no small correction
can patch the holes.

One can define an appropriate notion of a sparse set in the
framework of algorithmic randomness (Martin-L\"of definition of
randomness) by considering individual random sets (with respect to
the Bernoulli distribution $B_\varepsilon$) and their subsets as
``sparse.'' Then, we can prove that any sparse set (in this sense) satisfies
the conditions that are needed to make the iterative patching
procedure work. This algorithmic notion of ``sparseness'' is
discussed in~\cite{jac2008}. However, in the current paper we do not
assume that the reader is familiar with algorithmic randomness and
restrict ourselves to classical probability theory.

So our statements become quite lengthy and use probabilistic
quantifiers ``for almost all'' (= with probability $1$). The
order of quantifiers (existential, universal, and probabilistic)
is important here. For example, the statement ``a tile set
$\tau$ is robust'' means that \emph{there exists}
some~$\varepsilon>0$ such that \emph{for almost all~$E$} (with
probability~$1$ with respect to the distribution where each
point independently belongs to $E$ with probability
$\varepsilon$) the following is true: \emph{For every}
$(\tau,E)$-tiling $U$ \emph{there exists} a $\tau$-tiling $U'$
(of the entire plane) that is ``close'' to $U$. Here
by $(\tau,E)$-tiling we mean a tiling of $\mathbb{Z}^2\setminus E$
(where existing pairs of neighbor tiles match).

\subsection{Islands of errors}
\label{islands}

In this section we develop the notion of ``sparsity'' based on the
iterative grouping of errors (or holes) and prove its
properties.

Let $E\subset\mathbb{Z}^2$ be a set of points; points in $E$ are
called \emph{dirty}; other points are \emph{clean}. Let
$\beta\ge\alpha>0$ be integers. A nonempty set $X\subset E$ is
an \emph{$(\alpha,\beta)$-island} in~$E$ if

(1)~the diameter of $X$ does not exceed~$\alpha$ and

(2)~in  the $\beta$-neighborhood of $X$ there is no other point
from $E$.

(The diameter of a set is a maximal distance between its elements;
the distance $d$ is defined as $l_\infty$, i.e.,
the maximum of distances along
both coordinates; the $\beta$-neighborhood of $X$ is a set of all
points $y$ such that $d(y,x)\le\beta$ for some $x\in X$.)

It is easy to see that two (different) islands are disjoint (and
the distance between their points is greater than $\beta$).

Let $(\alpha_1,\beta_1), (\alpha_2,\beta_2),\ldots$ be a
sequence of pairs of integers and $\alpha_i\le \beta_i$ for all
$i$. Consider the following iterative ``cleaning'' procedure. At
the first step we find all $(\alpha_1,\beta_1)$-islands
(\emph{rank~1 islands}) and remove all their elements from $E$
(thus getting a smaller set $E_1$). Then, we find all
$(\alpha_2,\beta_2)$-islands in~$E_1$ (\emph{rank 2 islands});
removing them, we get $E_2\subset E_1$, etc. The cleaning process is
\emph{successful} if every dirty point is removed at some stage.

At the $i$th step we also keep track of the
$\beta_i$-neighborhoods of islands deleted during this step. A
point $x\in\mathbb{Z}^2$ is \emph{affected} during the $i$th step if
$x$ belongs to one of these neighborhoods.

The set $E$ is called \emph{sparse} (for a given sequence
$\alpha_i,\beta_i$) if the cleaning process is successful, and,
moreover, every point $x\in\mathbb{Z}^2$ is affected at finitely
many steps only (i.e., $x$ is far from islands of sufficiently
large ranks).

The values of $\alpha_i$ and $\beta_i$ should be chosen in such
a way that for sufficiently small $\varepsilon>0$ a
$B_\varepsilon$-random set is sparse with probability~$1$. (As we
have said, this justifies that our notion of sparsity is not
unreasonably restrictive.) The sufficient conditions are provided
by the following statement:
\begin{lem}\label{lemma1}
Assume that
        $$
8\sum_{k<n} \beta_k < \alpha_n \le \beta_n
\text{\quad for every $n$ and\quad}
\sum_i \frac{\log \beta_i}{2^i} <\infty.
        $$
Then, for all sufficiently small $\varepsilon>0$ a
$B_\varepsilon$-random set is sparse with probability~$1$.
\end{lem}

\begin{figure}
\begin{center}
\includegraphics[scale=1]{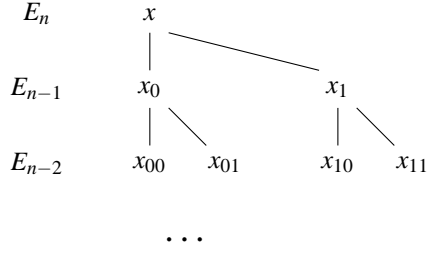}
\end{center}
\caption{Explanation tree; vertical lines connect different
names for the same points.}
\label{fpte7.mps}
\end{figure}

\begin{proof}
Let us estimate the probability of the event
``$x$ is not cleaned after $n$ steps'' for a given point $x$
(the probability of this event does not depend on $x$).
If $x\in E_n$, then $x$ belongs to $E_{n-1}$ and is not cleaned
during the $n$th step (when $(\alpha_n,\beta_n)$-islands in
$E_{n-1}$ are removed; by definition we let $E_0=E$). 
Then, $x\in E_{n-1}$ and, moreover, there
exists some other point $x_1\in E_{n-1}$ such that $d(x,x_1)$ is
greater than $\alpha_n/2$ but not greater than
$\beta_n+\alpha_n/2$ (note that $\beta_n+\alpha_n/2<2\beta_n$). Indeed, if there were no such
$x_1$ in $E_{n-1}$, then the $(\alpha_n/2)$-neighborhood of $x$ in
$E_{n-1}$ would be an $(\alpha_n,\beta_n)$-island in $E_{n-1}$ and $x$
would be removed.

Further, we apply the same argument on level $(n-1)$.
Each of the points $x_1$ and $x$ (we use notation $x_0$ for $x$, 
to make the notation more uniform) 
belongs to $E_{n-1}$; therefore it
belongs to $E_{n-2}$ together with some other point (at a
distance greater than $\alpha_{n-1}/2$ but not exceeding
$2\beta_{n-1}$). Denote these two other points in $E_{n-2}$ by
$x_{01}$ (which exists because $x_0\in E_{n-1}$) and $x_{11}$   
(which exists because $x_1\in E_{n-1}$) respectively. 
Thus, we have at least four points denoted by $x_{00}=x_0=x$,
$x_{01}$, $x_{10}=x_1$, and $x_{11}$ in $E_{n-2}$.
Then, we repeat the same argument for levels $(n-2), (n-3),$ etc.
In this way we get a tree
(Fig.~\ref{fpte7.mps}) that ``explains'' why $x$ belongs to $E_n$.

The distance between $x_0$ and $x_1$ in this tree is at least
$\alpha_n/2$ whereas the diameter of the subtrees starting at
$x_0$ and $x_1$ does not exceed
        $
\sum_{i<n} 2\beta_i.
        $
Therefore, the lemma's assumption guarantees that these subtrees
cannot intersect. Since it is true on all levels, all the leaves of the tree
are different. Note that all $2^n$ leaves of the tree belong to
$E=E_0$. As every point appears in $E$ independently of other
points, each ``explanation tree'' is valid with probability
$\varepsilon^{2^n}$. It remains to estimate the number of possible
explanation trees for a given point~$x$.

To specify $x_1$ we need to specify the difference (vertical and horizontal distances) 
between $x_0$ and $x_1$. Neither distance  exceeds
$2\beta_n$; therefore we need about $2\log (4\beta_n)$ bits to
specify them (including the sign bits). Then, we need to specify
the difference between $x_{00}$ and $x_{01}$ as well as the difference
between $x_{10}$ and $x_{11}$; this requires at most
$4\log(4\beta_{n-1})$ bits. To specify the entire tree we
therefore need
        $$
2\log (4\beta_n)+ 4 \log(4\beta_{n-1})+ 8 \log (4\beta_{n-2})+\cdots+
2^n \log (4\beta_1)
        $$
bits. Reversing the sum and taking out the factor $2^n$, 
we can rewrite this expression as
        $$
2^n (\log (4\beta_1)+  \log(4\beta_2)/2+\cdots).
        $$
Since the series $\sum \log \beta_n/2^n$ converges by
assumption, the total number of explanation trees for a given
point (and given $n$) does not exceed $2^{O(2^n)}$, so the
probability for a given point $x$ to be in $E_n$ for a
$B_\varepsilon$-random $E$ does not exceed
        $
\varepsilon^{2^n} 2^{O(2^n)},
        $
which tends to $0$ (even super-exponentially fast)
as~$n\to\infty$, assuming that $\varepsilon$ is small enough.

We conclude that the event ``$x$ is not cleaned'' (for a given
point $x$) has zero probability; the countable additivity
guarantees that with probability $1$ all points in
$\mathbb{Z}^2$ are cleaned.

It remains to show that every point with probability $1$ is
affected at finitely many steps only. Indeed, if $x$ is affected
at step $n$, then some point in its $\beta_n$-neighborhood
belongs to $E_n$, and the probability of
this event is at most
        $$
O(\beta_n^2)\varepsilon^{2^n} 2^{O(2^n)}=
2^{2\log \beta_n +O(2^n)-\log(1/\varepsilon)2^n};
        $$
the convergence conditions guarantees that $\log\beta_n=o(2^n)$,
so the first term is negligible compared to the others, the
probability series converges (for small enough~$\varepsilon$)
and the Borel--Cantelli lemma gives the desired result. \end{proof}

\smallskip

For our next step, we note that  by definition a sparse set is split into
a union of islands of different ranks. Now we prove that
these islands together occupy
only a small part of the plane. To formalize this statement,
we use the notion of Besicovitch size (density) of a set
$E\subset \mathbb{Z}^2$. Let us recall the definition. Fix some
point $O$ of the plane and consider squares of increasing size
centered at~$O$. For each square consider the fraction of points
in this square that belong to $E$. The $\limsup$ of these
frequencies is called the \emph{Besicovitch density} of $E$. (Note
that the choice of the center point~$O$ does not matter, since
for any two points $O_1$ and $O_2$ large squares of the same
size centered at
$O_1$ and $O_2$ share most of their points.)

\begin{figure}[h]
\begin{center}
\includegraphics{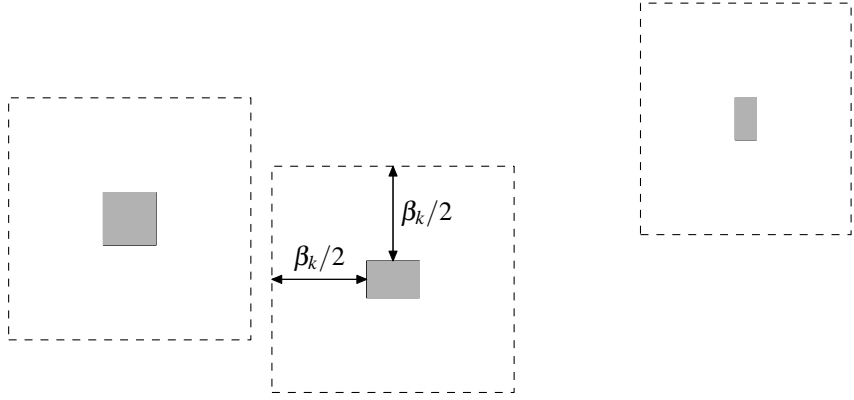}
\end{center}
\caption{Rank $k$ islands form a set of low density.
(In this picture each island is shown as a rectangle, which is
not always the case.)}\label{fig-8}
\end{figure}

By definition the distance between two rank $k$ islands is at
least $\beta_k$. Therefore the $(\beta_k/2)$-neighborhoods of these
islands are disjoint. Each of the islands contains at most
$\alpha_k^2$ points (it can be placed in a rectangle that has
sides at most $\alpha_k$). Each neighborhood has at least
$\beta_k^2$ points (since it contains a
$\beta_k\times\beta_k$ square centered at any point of the
island). Therefore the union of all rank $k$ islands has
Besicovitch density at most $(\alpha_k/\beta_k)^2$. Indeed, for
a large square the islands near its border can be ignored, and
all other islands are surrounded by disjoint neighborhoods where
their density is bounded by $(\alpha_k/\beta_k)^2$, see Fig.~\ref{fig-8}.

\begin{figure}[h]
\begin{center}
\includegraphics[scale=1]{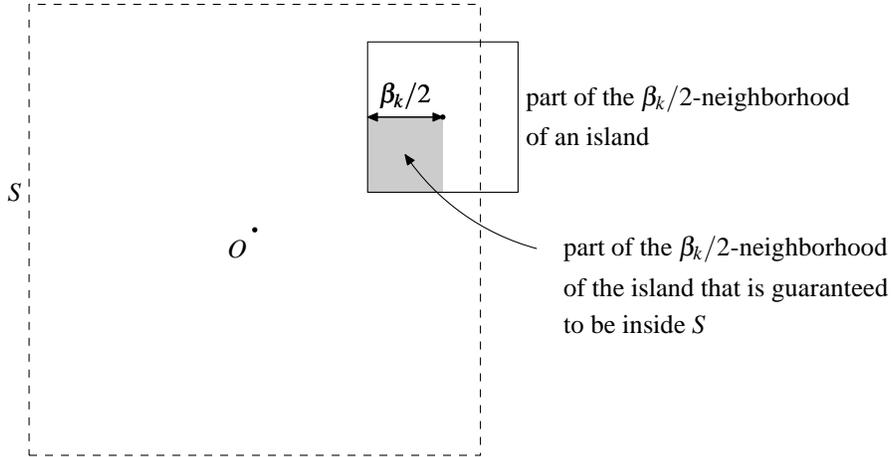}
\end{center}
\caption{Together with a point in a rank $k$ island, every square $S$
contains at least $(\beta_k/2)^2$ points of its
$(\beta_k/2)$-neighborhood.}\label{fig-9}
\end{figure}

One would like to conclude that the overall density of all
islands (of all ranks) does not exceed $\sum_k
(\alpha_k/\beta_k)^2$. However, the Besicovitch density is in
general not countably semiadditive (for example, the union of finite sets
having density $0$ may have density $1$), but in our case we are helped by the
second requirement of the definition of a sparse set (each point
is covered by only finitely many neighborhoods of islands).
\begin{lem}\label{lemma2}
Let $E$ be a sparse set for a given family of
$\alpha_k$ and $\beta_k$. Then, the Besicovitch density of $E$ is
$O(\sum (\alpha_k/\beta_k)^2)$.
\end{lem}
\begin{proof}
Let $O$ be a  center point used in
the definition of Besicovitch density. By definition of sparsity, this point is not covered by $\beta_k$-neighborhoods of
rank $k$ islands if $k$ is greater than some $K$. Now we split
the set $E$ into two parts: one ($E_\le$)  formed by islands
of rank at most $K$ and the other ($E_>$) formed by all islands of bigger ranks. As we have just seen,
in a large square the share of $E_\le$ is
bounded by $\sum_{k\le K} (\alpha_k/\beta_k)^2$ up to negligible
(as the size goes to infinity) boundary effects
(where we consider 
each $k\le K$ separately and then sum over all $k\le K$). A similar
bound is valid for rank $k$
islands with $k>K$, though the argument
is different and a constant factor appears.
Indeed, the $\beta_k$-neighborhood of every island $I$ 
does not contain the center point
$O$. Therefore, any square $S$ centered at $O$ that
intersects the island also contains  a significant part of its
$(\beta_k/2)$-neighborhood $N$: The intersection of $N$ and $S$
contains at least $(\beta_k/2)^2$ elements, see Fig.~\ref{fig-9}.
Therefore, the
share of $E_>$ in $S$ is bounded by $4\sum_{k>K}
(\alpha_k/\beta_k)^2$.
        \end{proof}

\textbf{Remark}. It is easy to choose $\alpha_k$ and $\beta_k$
satisfying the conditions of Lemma~\ref{lemma1}
and having arbitrarily
small $\sum(\alpha_k/\beta_k)^2$ (by taking geometric sequences that
grow fast enough). Therefore we get the following well-known
result as a corollary of Lemmas~\ref{lemma1} and~\ref{lemma2}:
For every $\alpha>0$
there exists $\varepsilon>0$ such that with probability $1$ a
$B_\varepsilon$-random set has Besicovitch density less than
$\alpha$. (In fact, a much stronger result is well known:
By the strong law of large numbers a $B_\varepsilon$-random
set has Besicovitch density $\varepsilon$ with probability~$1$.)
\medskip

In fact we will need a slightly more complicated version of
Lemma~\ref{lemma2}.
We are interested not only in the Besicovitch density
of a sparse set $E$ but also in the Besicovitch density of a
larger set: the union of $\gamma_k$-neighborhoods of rank $k$
islands in $E$. Here $\gamma_k$ are  some parameters;  in most applications
we set $\gamma_k=c\alpha_k$ for some constant $c$. The same argument
gives the bound $4\sum ((\alpha_k+2\gamma_k)/\beta_k)^2$.
Assuming that $\gamma_k\ge \alpha_k$, we can rewrite this bound
as $O(\sum (\gamma_k/\beta_k)^2)$. So we arrive at the
following statement:

\begin{lem}\label{lemma2a}
Let $E$ be a sparse set for a given family of $\alpha_k$ and $\beta_k$ and
let $\gamma_k\ge\alpha_k$ be some integers. Then, the union of $\gamma_k$-neighborhoods of level $k$ islands \textup(over all $k$ and all islands\textup)
has Besicovitch density $O(\sum(\gamma_k/\beta_k)^2)$.
\end{lem}

\subsection{Islands as a tool in percolation theory}
\label{percolation}

Let us show how some basic results of percolation theory can be
proved using the island technique.

\begin{thm}
For some $\alpha_k$ and $\beta_k$ satisfying the requirements of Lemma~\ref{lemma1}
the complement of any sparse set $E$ contains exactly one infinite
connected component $C$; the complement of $C$ has
Besicovitch density $O(\sum\alpha_k/\beta_k)^2$.
\end{thm}

\begin{proof} Let $\gamma_k=2\alpha_k$. (The choice of $\alpha_k$
and $\beta_k$ will be discussed later.) For every $k$ and for every
rank $k$ island fix a point in this island and consider the
$\gamma_k$-neighborhood of this point. It is a square containing
the entire island plus an additional ``security zone'' of width $\alpha_k$, 
contained in the $\gamma_k$-neighborhood of the island, see Fig.~\ref{fig-10}.

\begin{figure}[h]
\begin{center}
\includegraphics{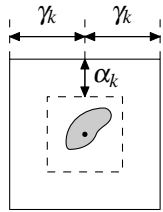}
\end{center}
\caption{A point in a rank $k$ island, its $\gamma_k$-neighborhood, and
the security zone of width $\alpha_k$.}\label{fig-10}
\end{figure}

It is enough to prove the following three statements:

\begin{itemize}
\item\emph{The union $U$ of all these squares
\textup(for all ranks\textup) contains the set~$E$ and
has Besicovitch density $O(\sum(\alpha_k/\beta_k)^2)$.}
\item\emph{The complement of $U$ is connected}.
\item\emph{There are no other infinite connected component in the
complements of $E$}.
\end{itemize}

The first statement is a direct corollary of Lemma~\ref{lemma2a}
above.

To prove the second statement, consider two points $x$ and $y$
outside $U$. We need to prove that $x$ and $y$ can be connected
by a path that is entirely outside $U$. Let us connect $x$ and $y$
by some path (say, one of the shortest paths) and then push this
path out of $U$.  Consider squares of maximal rank that intersect
this path. For each of them, consider the first moment when the
path gets into the square and the last moment when the path goes
out, and connect these two points by a path outside the square,
see Fig.~\ref{fig-11}.

\begin{figure}[h]
\begin{center}
\includegraphics{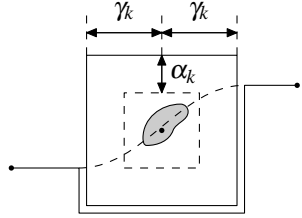}
\end{center}
\caption{Pushing a path out of the square.} \label{fig-11}
\end{figure}

Now the new path is $\alpha_k$-separated from this island of  rank $k$.
Provided $\beta_k - \gamma_k>\alpha_k$, the path after the correction
is $\alpha_k$-separated from all other islands of  rank greater than $k-1$. 
Indeed, the ``modified'' part of the corrected path (the points of the path involved into
 the correction procedure) remains at a distance of at most
$\gamma_k$ from the given $k$ level island; hence, it must remain at a distance 
at least $\alpha_k$ from all other islands of rank $k$ and higher. Note also that the shift (the distance between the original path and the
corrected one) does not exceed $2\gamma_k$.

Then, we can do the same for islands of rank $k-1$ (pushing
the path out of surrounding squares). Note that since at each step
the drift is bounded by $2\gamma_{k-1}$, we will not bump into
islands of rank $k$.

Repeating this process for decreasing $k$, we finally get a path
that connects $x$ and $y$ and goes entirely outside $U$. For
this we need only  the total drift on the smaller levels
(which is bounded by  $2\sum_{i<k} \gamma_i$) to be less than $\alpha_k$.
This is easy to achieve if $\alpha_k$, $\beta_k$, and $\gamma_k$
are suitable geometric sequences.

It remains to show that every infinite connected set intersects
the complement of $U$. To show this, let us take a big circular path
centered at the origin and
then push it out of $U$ as described above. Since the center
is outside the $\beta_k$-neighborhoods of islands for large enough $k$,
we may assume that the sizes of islands that intersect this circle
are small compared to its radius (say, less than $1\%$ of it, which
can be guaranteed if the geometric sequences $\alpha_k$,
$\beta_k$, and $\gamma_k$ grow fast enough).  Then, after the change
the circle will still encircle a large neighborhood of the origin, so
any infinite connected component should cross such a circle.
\end{proof}

\subsection{Bi-islands of errors}
\label{bi-islands}

In the proof of our main result (Section~\ref{robust-complex}) we
need a more delicate version of the definition of islands.
In fact we need such a definition that some
counterpart of  Lemma~\ref{lemma1} could be applied
even if the sequence $\log \beta_n$ grows
much faster than $2^n$ (e.g., for $\beta_n=c^{(2.5)^n}$).
In this section we  define bi-islands (a generalization of the notion
of islands from Section~\ref{islands}) and prove
bi-island versions of Lemmas~\ref{lemma1},~\ref{lemma2}, and~\ref{lemma2a}.
The reader can safely skip this section for now
and return here before reading Section~\ref{robust-complex}.

Let $E\subset\mathbb{Z}^2$ be a set of points. As in
Section~\ref{islands}, we call points in $E$ \emph{dirty}, and the
other points \emph{clean}.
Let $\beta\ge\alpha>0$ be integers. A nonempty set $X\subset E$ is
an \emph{$(\alpha,\beta)$-bi-island} in~$E$ if $X$ can be represented
as the union of some sets $X_0$, $X_1$ such that

(1)~in the $\beta$-neighborhood of $X=X_0\cup X_1$ there are no points
from $E\setminus X$; 

(2)~the diameters of $X_0$ and $X_1$ do not exceed~$\alpha$; and

(3)~the distance between $X_0$ and $X_1$ does not exceed~$\beta$.

\begin{figure}[h]
\begin{center}
\includegraphics[scale=0.8]{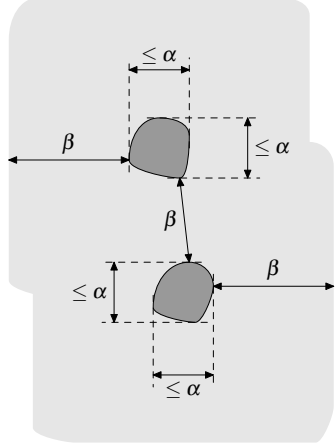}
\end{center}
\caption{A bi-island, a union of two ``islands'' that are close to each other.}
\label{biisland}
\end{figure}

\noindent
(See Fig.~\ref{biisland}.)
In particular, an $(\alpha,\beta)$-island is a special case
of an $(\alpha,\beta)$-bi-island (by letting $X_1$ be empty).

Note that one may split the same bi-island into $X_0$ and $X_1$ in different ways.

Obviously, every two different bi-islands are disjoint. Moreover,
the distance between them is greater than $\beta$.
The diameter of a bi-island is at most
$(2\alpha+\beta)$.

Let $(\alpha_1,\beta_1), (\alpha_2,\beta_2),\ldots$ be a
sequence of pairs of integers and $\alpha_i\le \beta_i$ for all
$i$. We define an iterative cleaning procedure for bi-islands. At
the first step we find all $(\alpha_1,\beta_1)$-bi-islands
and remove all their elements from $E$ (getting a smaller set $E_1$).
Then, we find in $E_1$ all $(\alpha_2,\beta_2)$-bi-islands;
removing them, we get $E_2\subset E_1$, etc. The cleaning process is
\emph{successful} if every dirty point is removed at some stage.

Similarly to the case of islands, we say that a
point $x\in\mathbb{Z}^2$ is \emph{affected} during step $i$ if
$x$ belongs to the $\beta_i$-neighborhood of one of the bi-islands of
rank $i$.

The set $E$ is called \emph{bi-sparse} (for a given sequence
$\alpha_i,\beta_i$) if the cleaning process defined above is successful,
and, moreover, every point $x\in\mathbb{Z}^2$ is affected at finitely
many steps only (which means that $x$ is far from bi-islands of
sufficiently large ranks).

We choose the values of $\alpha_i$ and $\beta_i$
in such a way that for sufficiently small $\varepsilon>0$ a
$B_\varepsilon$-random set is bi-sparse with probability~$1$.
The main achievement here is that the convergence condition is now weaker
than in the corresponding statement for islands (Lemma~\ref{lemma1}):

\begin{lem}
\label{lemma-bi-island}
Assume that
        $$
12\sum_{k<n} \beta_k < \alpha_n \le \beta_n
\text{\ for every $n$,  and\quad}
\sum_i \frac{\log \beta_i}{3^i} <\infty.
        $$
Then, for all sufficiently small $\varepsilon>0$, a
$B_\varepsilon$-random set is bi-sparse with probability~$1$.
\end{lem}

\begin{proof} The proof of Lemma~\ref{lemma-bi-island}
is very similar to the proof of
Lemma~\ref{lemma1}. At first we estimate the probability of the event
``$x$ is not cleaned after $n$ steps'' for a given point $x$.
If $x\in E_n$, then $x$ belongs to $E_{n-1}$ and is not cleaned
during the $n$th step (when $(\alpha_n,\beta_n)$-bi-islands in
$E_{n-1}$ are removed). Then, $x\in E_{n-1}$. Moreover, we
show that there exist \emph{two other} points $x_1,x_2\in E_{n-1}$
such that the three distances $d(x,x_1)$, $d(x,x_2)$, and $d(x_1,x_2)$
are all greater than $\alpha_n/2$ but not greater than
$2\beta_n+2(\alpha_n/2) < 3\beta_n$.

Let $X_0$ be the $(\alpha_n/2)$-neighborhood of $x$ in $E$. If $X_0$ were
an island, it would be removed. Since this does not occur, there is a point
$x_1$ outside $X_0$ but in the $\beta_n$-neighborhood of $X_0$.

Let $X_1$ be the $(\alpha_n/2)$-neighborhood of $x_1$ in $E$. Again $X_0$ and $X_1$ do not form a bi-island.
Both sets $X_0$ and $X_1$ have diameter at most $\alpha_n$, and
the distance between them is at most $\beta_n$. So the only reason why they
are not a bi-island is that there exists a point $x_2\in E$ outside $X_0\cup X_1$
but in the $\beta_n$-neighborhood of it. The points $x_1$ and $x_2$ have the
required properties (the distances $d(x,x_1)$, $d(x,x_2)$, and $d(x_1,x_2)$
are  greater than $\alpha_n/2$ but not greater than $3\beta_n$).

To make the notation uniform, we denote $x$ by $x_0$.
Each of the points $x_0, x_1, x_2$ belongs to $E_{n-1}$. This
means that each of them belongs to $E_{n-2}$ together with
a pair of other points (at a distance greater than
$\alpha_{n-1}/2$ but not exceeding $3\beta_{n-1}$). In this way
we get a ternary tree that ``explains'' why $x$ belongs to $E_n$.

The distance between every two points among $x_0$, $x_1$, and $x_2$
in this tree is at least $\alpha_n/2$ whereas the diameters of the
subtrees starting at $x_0$, $x_1$, and $x_2$ do not exceed
        $
\sum_{i<n} 3\beta_i.
        $
Thus, the lemma's assumption guarantees that these subtrees
cannot intersect and that all the leaves of the tree
are different. The number of leaves in this ternary tree is $3^n$,
and they all belong to $E=E_0$. Every point appears in $E$ independently
of other points; hence, one such  ``explanation tree'' is valid with
probability $\varepsilon^{3^n}$. It remains to count the number of all
explanation trees for a given point~$x$.

To specify $x_1$ and $x_2$ we need to specify horizontal and vertical
distances between $x_0$ and $x_1, x_2$. These distances do not exceed
$3\beta_n$; therefore we need about $4\log (6\beta_n)$ bits to
specify them (including the sign bits). Then, we need to specify
the distances between $x_{00}$ and $x_{01}, x_{02}$ as well as the
distances between $x_{10}$ and $x_{11}, x_{12}$ and between $x_{20}$ and
$x_{21}, x_{22}$. This requires at most
$12\log(6\beta_{n-1})$ bits. To specify the entire tree we
therefore need
        $$
4\log (6\beta_n)+ 12 \log(6\beta_{n-1})+ 36 \log (6\beta_{n-2})+\cdots+
4\cdot 3^{n-1} \log (6\beta_1),
        $$
which is equal to
        $
4\cdot 3^{n-1} (\log (6\beta_1)+  \log(6\beta_2)/3+\cdots).
        $
The series $\sum \log \beta_n/3^n$ converges by
assumption; so, the total number of explanation trees for a given
point (and given $n$) does not exceed $2^{O(3^n)}$. Hence, the
probability for a given point $x$ to be in $E_n$ for a
$B_\varepsilon$-random $E$ does not exceed
        $
\varepsilon^{3^n} 2^{O(3^n)},
        $
which tends to $0$ as~$n\to\infty$
(assuming that $\varepsilon$ is small enough).

We conclude that the event ``$x$ is not cleaned'' (for a given
point $x$) has zero probability; hence,
with probability $1$ \emph{all} points in $\mathbb{Z}^2$ are cleaned.

It remains to show that every point with probability $1$ is
affected by finitely many steps only. Indeed, if $x$ is affected
by step $n$, then some point in its $\beta_n$-neighborhood
belongs to $E_n$, and the probability of
this event is at most
        $$
O(\beta_n^2)\varepsilon^{3^n} 2^{O(3^n)}=
2^{2\log \beta_n +O(3^n)-\log(1/\varepsilon)3^n}.
        $$
From the convergence conditions we have $\log\beta_n=o(3^n)$,
so the first term is negligible compared to others. The
probability series converges (for small enough~$\varepsilon$)
and the Borel--Cantelli lemma gives the  result. \end{proof}

\smallskip

By definition, a bi-sparse set is split into a union of bi-islands of
different ranks. Such bi-islands occupy only a small part of the plane:
\begin{lem}\label{lemma-bi-island-2}
Let $E$ be a bi-sparse set for a given
family of $\alpha_k$ and $\beta_k$. Then, the Besicovitch density of $E$
is $O(\sum (\alpha_k/\beta_k)^2)$.
\end{lem}

\begin{proof} The proof of Lemma~\ref{lemma-bi-island-2} repeats the proofs of
Lemma~\ref{lemma2}.\end{proof}

Recalling Lemma~\ref{lemma2a}, we may consider
a sequence of numbers  $\gamma_k$ such that
$\gamma_k\ge \alpha_k$. Then, the Besicovitch density of the
union of $\gamma_k$-neighborhoods
of rank $k$ bi-islands (for all $k$ and for all islands) is bounded by $O(\sum (\gamma_k/\beta_k)^2)$.

However, this statement is not enough for us.
In Section~\ref{robust-complex} we will need a kind of  ``closure''
of the  $\gamma_k$-neighborhood of a bi-island:

\begin{definition}
Let $S$ be a $k$-level bi-island. We say that $(x,y)\in\mathbb{Z}^2$
belongs to the \emph{extended} $\gamma$-neighborhood of $S$
if  there exist two points $(x,y'), (x,y'')\in\mathbb{Z}^2$ \textup(with the same first coordinate\textup)
such that
$\mathrm{dist}(S, (x,y'))\le \gamma$,
$\mathrm{dist}(S, (x,y''))\le \gamma$, and $y'\le y \le y''$ (see Fig.~\ref{biisland-neighb}).
\begin{figure}[h]
\begin{center}
\includegraphics[scale=0.8]{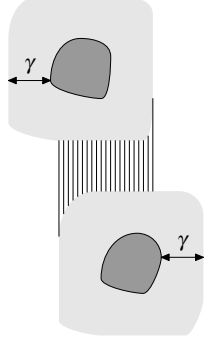}
\end{center}
\caption{An extended neighborhood of a bi-island consists of the neighborhoods of its two parts and a zone between them.}
\label{biisland-neighb}
\end{figure}
\end{definition}
The meaning of the last definition is quite simple: We take not
only the points that are close to $S$ but also those points that
are placed somehow between the neighborhoods of
$S_0$ and $S_1$.
\begin{lem}\label{lemma-bi-island-3}
Let $E$ be a bi-sparse set for a given
family of $\alpha_k$ and $\beta_k$ satisfying the conditions
of Lemma~\ref{lemma-bi-island}.  Let $\gamma_k$ be a
sequence of numbers  such that $\alpha_k <\gamma_k$, and
the series $\sum(\gamma_k/\beta_k)$ converges.
Then, the Besicovitch density of
the union of extended $\gamma_k$-neighborhoods
of rank $k$ bi-islands in $E$ is bounded by $O(\sum (\gamma_k/\beta_k))$.
\end{lem}

\begin{proof} The arguments are
similar to the proof of  Lemma~\ref{lemma2a}.
An extended $\gamma_k$-neighborhood of a $k$-level island can
be covered by a rectangle of width $O(\gamma_k)$ and height
$O(\beta_k+\gamma_k)$; so its area is
$O(\gamma_k\beta_k)$ (since $\gamma_k\le \beta_k$).
The distance between any two bi-islands of rank
$k$ is at least $\beta_k$.
Hence, the fraction of \emph{extended} $\gamma_k$-neighborhoods
of islands
is $O(\sum \gamma_k/\beta_k)$ (this is similar to the bound
$O(\sum (\gamma_k/\beta_k)^2)$, which holds for simple
$\gamma_n$-neighborhoods).~\end{proof}

\medskip

 Lemmas~\ref{lemma-bi-island}--\ref{lemma-bi-island-3} will be used in Section~\ref{robust-complex}.
(The arguments of Sections~\ref{robust}--\ref{strongly-robust} do not refer to bi-islands.) These lemmas will be used
for $\alpha_k,\beta_k$ such that $\log \alpha_k\sim q^k$
for $q>2$,  $\beta_k\sim \alpha_{k+1}$, and
$\gamma_k = O(\alpha_k)$ or $\gamma_k=O(\alpha_k^2)$.
Note that  we cannot apply Lemmas~\ref{lemma1} and~\ref{lemma2}
(about \emph{islands}) for these parameters because $\log \beta_k$ grows faster than $2^k$.
So we need to deal with bi-islands.

In the definition of sparse sets
in Section~\ref{islands} each single island of rank $k$ must be isolated from other islands
of rank $k$. In this section we modified this definition and allowed  an island to be close to
at most one other island of the same rank.
In a similar way, we could define $s$-islands for any $s\ge 2$, assuming that clusters of $s$
islands of rank $k$ (rather close to each other) are authorized.
A set that can be represented as a union of $s$-islands of different ranks can be called $s$-sparse.
A generalization of Lemmas~\ref{lemma-bi-island} can be proven:
A random set is $s$-sparse with probability $1$ if $\sum (\log \beta_i)/(s+1)^i$ converges.
However, we do not develop here the general theory of $s$-sparse sets.
The concept of bi-islands and bi-sparsity (i.e., the case $s=2$) is enough for all our applications
in Section~\ref{robust-complex}.

\section{Robust tile sets}
\label{robust}

In this section we construct an aperiodic
\begin{figure}[h]
\begin{center}
        $$
\includegraphics[scale=0.7]{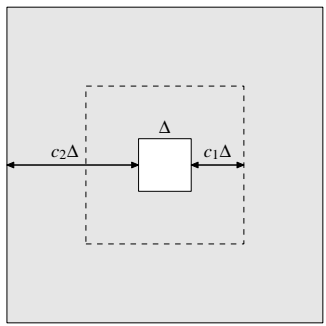}
        $$
\end{center}
\caption{Patching holes.}
\label{fpt.5.mps}
\end{figure}
tile set where isolated defects can be healed.

Let $c_1<c_2$ be positive integers. We say that a tile set
$\tau$ is \emph{$(c_1,c_2)$-ro\-bust} if the following holds: For
every $\Delta$ and for every $\tau$-tiling $U$ of the
$(c_2\Delta)$-neighborhood of a square $\Delta\times \Delta$ excluding the square
itself there exists a tiling $V$ of the entire
$(c_2\Delta)$-neighborhood of the square (including the square itself)
that coincides with $U$ outside of the $(c_1\Delta)$-neighborhood of
the square (see Fig.~\ref{fpt.5.mps}).

\begin{thm}
        \label{robust-tileset}
There exists a self-similar tile set that is $(c_1,c_2)$-robust
for some $c_1$ and $c_2$.
\end{thm}

\begin{proof} For every tile set $\mu$ it is easy to construct
a ``robustified'' version $\mu'$ of $\mu$, i.e., a tile set
$\mu'$ and a mapping $\delta\colon\mu'\to\mu$ such that
(a)~$\delta$-images of $\mu'$-tilings are exactly $\mu$-tilings and
(b)~$\mu'$ is ``5-robust'': Every $\mu'$-tiling of a $5\times 5$
square minus $3\times 3$ hole (see Fig.~\ref{fpt.6.mps}) can be
uniquely extended to the tiling of the entire $5\times 5$
square.

\begin{figure}[h]
\begin{center}
        $$
\includegraphics[scale=1.0]{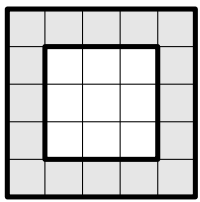}
        $$
\end{center}
\caption{Filling a $3\times 3$ hole.}
\label{fpt.6.mps}
\end{figure}

Indeed, it is enough to keep in one $\mu'$-tile the information
about the $5\times 5$ square in $\mu$-tiling. Matching rules
will guarantee that the information about the intersection
($4\times 5$ rectangle) is consistent in neighbor tiles.
Then, a $3\times 3$ hole (as shown in the picture) is not
fatal. It is easy to see that the tiles at its border (gray) are consistent
and contain all the information the missing tiles should have.
 (In fact,  using more careful estimates one can replace in our argument
 the $5\times5$ squares  by $4\times4$ squares; but we do not care much
 about constants.)

This robustification can be easily combined with the fixed-point
construction. In this way we can get a ``$5$-robust'' self-similar
tile set $\tau$ if the zoom factor $N$ (which is considered to be fixed
in this argument) is large enough. It remains to explain that ``$5$-robustness''
(in the sense described above) implies also $(c_1,c_2)$-robust for some $c_1$ and
$c_2$. (The values of $c_1$ and $c_2$ depend on $N$, but $N$ is fixed.)

Indeed, assume that a tiling of a large enough neighborhood
around a $\Delta\times \Delta$ hole is given. Denote by $k$ the minimal
integer such that $N^k\ge \Delta$ (so the $k$-level macro-tiles are
greater than the hole under consideration). Note that the size
of the $k$-level macro-tiles is linear in $O(\Delta)$ since $N^k \le N\cdot
\Delta$.

In the tiling around the hole, an $N\times N$ block structure is
correct except for the $N$-neigh\-bor\-hood of the central $\Delta\times
\Delta$ hole. Indeed, the colors encode coordinates, so in every
connected tiled region the coordinates are consistent. For similar
reasons an $N^2\times N^2$ structure is correct except for the
$(N+N^2)$-neighborhood of the hole, etc. Hence, for the chosen $k$ we get a
$k$-level structure that is correct except for (at most)
$9=3\times 3$ squares of level $k$, so we can delete everything
in these squares and use $5$-robustness to replace them with macro-tiles
that correspond to replacement tiles.

To start this procedure (and fill the hole), we need a correct
tiling only in the $O(N^k)$ neigh\-bor\-hood of the hole.
(Technically, we need to have a correct tiling in the
$(3N^k)$-neighborhood of the hole; as $3N^k\le 3N\Delta$, we let
$c_2=3N.$) The correction procedure involves changes in another
$O(N^k)$-neighborhood of the hole. (Technically, the changes touch
$(2N^k)$-neighborhood of the hole; $2N^k\le 2N\Delta$, so we let
$c_1=2N$.)
\end{proof}

\section{Robust tile sets with variable zoom factors}
\label{robust-var}

The construction from the previous section works only for
self-similar tilings with a fixed zoom factor. It is enough for
simple applications, as we will see in
Section~\ref{strongly-robust}. However, in the proof of our main
result in Section~\ref{robust-complex} we need a variable
zoom factor. So here we develop a technique suitable for this
case. This section can be skipped now but it should
be read before Section~\ref{robust-complex}.

Now we explain how to get
``robust''  fixed-point tilings with variable zoom factors
$N_1, N_2,\ldots$.
As well as in the case of a fixed zoom factor,
the idea is that $k$-level macro-tiles are ``responsible''
for healing holes of size comparable with these macro-tiles.

Let $\Delta_0\le \Delta_1 \le \Delta_2 \le \ldots$ be a sequence of integers.
Let $c_1<c_2$ be positive integers. We say that a tile set
$\tau$ is \emph{$(c_1,c_2)$-ro\-bust against holes of size
$\Delta_0,\Delta_1,\ldots$}
if the following holds: For
every $n$ and for every $\tau$-tiling $U$ of the
$c_2\Delta_k$ neigh\-bor\-hood of a square $\Delta_k\times \Delta_k$
excluding the square itself there exists a tiling $V$ of the entire
$c_2\Delta_k$ neigh\-bor\-hood of the square (including the square itself)
that coincides with $U$ outside of the $c_1\Delta_k$ neigh\-bor\-hood of
the square.
The difference from the  definition of Section~\ref{robust} is that
we take only values $\Delta\in\{\Delta_0,\Delta_1,\ldots\}$
instead of holes of arbitrary size.

\begin{lem}
        \label{robust-tileset-var}
Assume a sequence of zoom factors $N_k$ grows not too fast
and not too slow \textup(it is enough to assume that $N_k \ge C\log k$
and $C \log N_{k+1} < N_k$ for a large enough $C$; cf. the discussion
in Section~\ref{variable}\textup).
Then, there exists a
tile set with variable zoom factors $N_k$ \textup($k$-level macro-tiles
of size $L_k=N_0\cdots  N_{k-1}$\textup)
that is  $(c_1,c_2)$-robust \textup(for some $c_1$ and $c_2$\textup) against
holes of size $L_0,L_1,\ldots$.
\end{lem}

\begin{proof} First, we apply the fixed-point construction
from Section~\ref{variable} and get a tile set
that is ``self-similar'' with variable zoom factors $N_1,N_2,\ldots$.
Denote by $\mu_k$ the family of $k$-level macro-tiles corresponding
to  this tile set.

Further we make a  ``robustified'' version of this tile set.
To this end we basically repeat the arguments from Section~\ref{robust}
(the proof of Theorem~\ref{robust-tileset}).  The difference in the argument
is  that now we deal with variable zoom factors, and sizes of holes are taken from the
sequence $L_0,L_1,\ldots$.

Denote by $\mu'_k$ the family
of  $k$-level macro-tiles for the new tiling. We need that
there exists a mapping $\delta\colon\mu'_k\to\mu_k$ such that
(a)~$\delta$-images of $\mu'_k$-tilings are exactly $\mu_k$-tilings and
(b)~$\mu'_k$ is ``5-robust'': Every $\mu'_k$-tiling of a $5\times 5$
square minus a $3\times 3$ hole (see  again Fig.~\ref{fpt.6.mps}) can be
uniquely extended to the tiling of the entire $5\times 5$
square.

To get such a robustification, it is enough to keep in every
$\mu'_k$-macro-tile the information about the  $5\times 5$ square
in the $\mu_k$-tiling and use the colors on the borders to ensure
that this information is coherent in neighbor macro-tiles.

As usual, this robustification can be combined with the fixed-point
construction. We get  $5$-robust macro-tiles for all levels of our
construction. ``Self-similarity'' guarantees that the same property holds
for macro-tiles of all levels, which implies the required property of
generalized robustness.

Indeed, assume that a tiling of a large enough neighborhood
around a $\Delta \times \Delta$ hole is given, and
$\Delta\le L_k$ for some $k$.
In the tiling around the hole, an $(L_1\times L_1)$
block structure, is correct except  for only the
$L_1$ neigh\-bor\-hood of the hole.
For similar reasons an $(L_2\times L_2)$ structure is
correct except for the $(L_1+L_2)$ neigh\-bor\-hood, etc.
So  we get a $k$-level structure that is correct except
for (at most) $9=3\times 3$ squares of size $L_k\times L_k$.
Because of $5$-robustness, this
hole  can be filled with $k$-level
macro-tiles. Note that reconstruction of ground-level tiles inside a high-level macro-tile is unique
after we know its ``conscious known'' information, i.e., the content of the tape of the
Turing machine simulated on the computation zone  of this macro-tile.
(This information is reconstructed from
the consciously known information of the neighbor macro-tiles.)
[For the maximal complexity tile set (Section~\ref{complex}) it is not
the case, and the absence of this property will become a
problem in Section~\ref{robust-complex}
where we robustify it. To solve this problem, we will need to
use error-correcting codes.]

To implement the patching procedure (and fill the hole) we need
to have a correct tiling  in the $O(L_k)$ neigh\-bor\-hood of the
hole. The correction procedure involves changes in another
$O(L_k)$ neigh\-bor\-hood of the hole.
More technically, we need to have a correct tiling in the
$(3L_k)$-neighborhood of a hole of size $L_k$, so we let
$c_2=3$. Since the correction procedure involves changes in the
$(2L_k)$-neighborhood of the hole, we let
$c_1=2$.
\end{proof}

\medskip

We can robustify tiling not only against holes but against
\emph{pairs of holes}. To this end we slightly modify our definition of
robustness. Let $\Delta_0\le \Delta_1 \le \Delta_2 \le \ldots$
be an increasing sequence of integers, and let $c_1<c_2$ be positive
integers. We say that a tile set $\tau$ is \emph{$(c_1,c_2)$-ro\-bust
against pairs of holes of size  $\Delta_0,\Delta_1,\ldots$}
if the following holds:
Let us have two sets $H_1, H_2 \subset \mathbb{Z}^2$, each of
them of diameter at most $\Delta_k$ (for some $k>0$). For every
$\tau$-tiling $U$ of  the $c_2\Delta_k$ neigh\-bor\-hood of the union
$(H_1\cup H_2)$ excluding $H_1$ and $H_2$ themselves
there exists a tiling $V$ of the entire $c_2\Delta_k$ neigh\-bor\-hood
of $(H_1\cup H_2)$ (including $H_1$ and $H_2$ themselves)
that coincides with $U$ outside of the $c_1\Delta_k$ neigh\-bor\-hood of
$(H_1\cup H_2)$.

A robustification against pairs of holes can be done in the same
way as the robustification against a single isolated hole. Indeed,
if these two holes are far apart from each other, we can ``correct''  them
independently;  if they are rather close to each other, we
correct them as one hole of (roughly) doubled size. So
we can employ the same robustification technique as before;
we need only to take a large enough
``radius of  multiplication'' $D$ (and use $D$-robustness instead
of $5$-robustness). So we get the following generalization of
Lemma~\ref{robust-tileset-var}:
\begin{lem}
        \label{robust-tileset-var-pairs}
Assume a sequence of zoom factors $N_k$ grows not too fast
and not too slow \textup(e.g., $N_k \ge C\log k$
and $C \log N_{k+1} < N_k$ for a large enough $C$\textup).
Then, there exists a
tile set with zoom factors $N_k$ (i.e., with $k$-level macro-tiles of size $L_k=N_0\cdots  N_{k-1}$\textup)
that is  $(c_1,c_2)$-robust \textup(for some $c_1$ and $c_2$\textup) against
pairs of holes of size $L_0,L_1,\ldots$ for some $c_1$ and $c_2$.
\end{lem}
Of course, similar propositions can be also proven for triplets, quadruplets,
and any other sets of holes of bounded cardinality.
However, in this paper we consider only pairs of holes; this is enough for our argument in
Section~\ref{robust-complex}.

\section{Strongly aperiodic robust tile sets}
\label{strongly-robust}

Now we are ready to apply the islands technique to construct a robust
strongly aperiodic tile set. We start with a formal definition of a tiling with errors
(see the motivation and discussion in Section~\ref{random}).
\begin{definition}
For a subset $E\subset \mathbb{Z}^2$ and a tile set
$\tau$ we call by  a $(\tau,E)$-tiling any mapping
 $$ T\ : \ (\mathbb{Z}^2\setminus E )\to \tau$$
such that for every two neighbor cells
$x,y\in \mathbb{Z}^2\setminus E$,  tiles $T(x)$
and $T(y)$ satisfy the tiling rules \textup(colors on adjacent
sides match\textup).
We may say that $T$ is a $\tau$-tiling of the plane
with holes at points of $E$.
 \end{definition}

\begin{thm}
        \label{thm:strongly-aperiodic-robust}
There exists a tile set $\tau$ with the following properties:
\textup{(1)}~$\tau$-tilings of $\mathbb{Z}^2$ exist and
\textup{(2)}~for all sufficiently small $\varepsilon$ for almost
every \textup(with respect to $B_\varepsilon$\textup) subset
$E\subset\mathbb{Z}^2$ every $(\tau,E)$-tiling is at least
$1/10$ Besicovitch apart from every periodic mapping
$F\ :\ \mathbb{Z}^2\to\tau$.
\end{thm}

\textbf{Remark 1}. Since the tiling contains holes, we need to
specify how we treat the holes when defining the Besicovitch
distance. We do \emph{not} count points in $E$ as points where
two mappings differ; this makes our statement stronger.

\textbf{Remark 2}. The constant $1/10$ is not optimal and can be
replaced by any other constant  $\alpha < 1$.

\begin{proof} Consider a tile set $\tau$ such that (a)~all
$\tau$-tilings are $\alpha$-aperiodic for every $\alpha<1/4$ and
(b)~$\tau$ is $(c_1,c_2)$-robust for some $c_1$ and $c_2$. Such
a tile set can be constructed by combining the arguments
used for Theorems~\ref{robust-tileset}  and
~\ref{aperiodic}. More precisely,
we take  as the ``basic'' construction
the tile set from the proof of Theorem~\ref{aperiodic}
(which simulates the Thue--Morse substitution).
Then, we ``robustify''  it by the procedure from the proof of Theorem~\ref{robust-tileset}. For the robustified tile set we know that each macro-tile
in a tiling keeps  the conscious information that was given (in the ``basic'' tile set)
to all macro-tiles in its $5\times5$-neighborhood; so the new tiling is
not only strongly aperiodic but also $5$-robust. It remains to show that this construction
implies claim~(2) of the theorem.

We want to apply our  probabilistic  lemmas concerning ``island of errors''.
We need to choose  $\alpha_k$ and $\beta_k$ such that 

\begin{itemize}

\item the conditions of Lemma~\ref{lemma1} (p.~\pageref{lemma1})
are satisfied, and therefore
a random error set with probability $1$ is sparse with respect to these $\alpha_k$
and $\beta_k$;

\item for every sparse set $E\subset\mathbb{Z}^2$, every
$(\tau,E)$-tiling can be iteratively corrected (by changing it
in the neighborhoods of islands of all ranks) into
a $\tau$-tiling of the entire plane;
and
\item the Besicovitch distance between the tilings before and
after correction is small.

\end{itemize}

Then, we conclude that the original $(\tau,E)$-tiling is strongly
aperiodic since the corrected tiling is strongly aperiodic and
close to the original one.

To implement this plan, we use the following lemma that
describes the error-correction process.
\begin{lem} \label{lemma-robust-1}
Assume that a tile set $\tau$ is
$(c_1,c_2)$-robust, $\beta_k>4c_2\alpha_k$ for every $k$, and a
set $E\subset \mathbb{Z}^2$ is sparse \textup(with parameters
$\alpha_k$, $\beta_k$\textup). Then, every $(\tau,E)$-tiling can be
transformed into a $\tau$-tiling of the entire plane by changing
it in the union of $(2c_1\alpha_k)$-neighborhoods of rank $k$
islands \textup(for all islands of all ranks).
\end{lem}
\begin{proof} Note that $(\beta_k/2)$-neighborhoods
of rank $k$ islands are disjoint and large enough to perform the
error correction of rank~$k$ islands, since
$\beta_k>4c_2\alpha_k$. The definition of a sparse set
guarantees also that every point is changed only finitely many times
(so the limit tiling is well defined) and that the limit tiling has no errors.\end{proof}

The Besicovitch density of the changed part of a tiling can be
estimated using Lemma~\ref{lemma2}.
Here $\gamma_k=2c_1\alpha_k$
is proportional to $\alpha_k$, so the
Besicovitch distance between the original
and corrected tilings (in Lemma~\ref{lemma-robust-1})
is $O(\sum_k (\alpha_k/\beta_k)^2)$.
(Note that the constant in $O$ notation depends on $c_1$.)

It remains to choose $\alpha_k$ and $\beta_k$. We have to satisfy
all the inequalities in Lemmas~\ref{lemma1}, \ref{lemma2},
and~\ref{lemma-robust-1}. To satisfy
Lemmas~\ref{lemma2} and \ref{lemma-robust-1},
we may let $\beta_k = ck\alpha_k$ for large
enough $c$. To satisfy Lemma~\ref{lemma1}, we may let
$\alpha_{k+1}=8(\beta_1+\cdots+\beta_k)+1$. Then, $\alpha_k$ and
$\beta_k$ grow faster than any geometric sequence (like
$k!$ multiplied by some exponent in $k$), but still $\log
\beta_k$ is bounded by a polynomial in $k$ and the series in
Lemma~\ref{lemma1} converges.

With these parameters (and taking $c$ large enough) we  guarantee
that the Besicovitch distance between the original $(\tau,E)$-tiling
and the corrected $\tau$-tiling does not exceed, say, $1/100$.

Now assume that some $(\tau,E)$-tiling $V$ is at a distance less than
$1/10$ from  some periodic configuration $W$ (with a period $v$). As we  
just explained, the original $(\tau,E)$-tiling $V$ must be at a distance 
at most $1/100$ from some correct $\tau$-tiling $V'$. Let us consider the $v$-shift of 
both  configurations $V$ and $V'$ ($W$ is shifted to itself). It is easy to see that
the distance between the initial and the shifted copies of configuration $V'$ 
is not greater
than the sum  $\mathrm{dist}(V',V)+\mathrm{dist}(V,W)$  
taken twice.  Since the corrected tiling $V'$ must be $1/4$-aperiodic, and
$1/4>2(1/10+1/100)$, we get a contradiction.\end{proof}

\section{Robust tile sets that enforce complex tilings}
\label{robust-complex}

In this section we prove the main result of the paper. We construct a tile
set that guarantees large Kolmogorov complexity of every tiling and
that is robust with respect to random errors.
\begin{thm}\label{thm-complex-errors}
There exists a tile set $\tau$ and constants $c_1,c_2>0$
with the following properties:

\textup{(1)} a $\tau$-tiling of  $\mathbb{Z}^2$ exists;

\textup{(2)} for every $\tau$-tiling~$T$ of the plane,
every $N\times N$ square
of $T$ has Kolmogorov complexity  at least $c_1N-C_2$;

\textup{(3)} for all sufficiently small $\varepsilon$ for almost every
\textup(with respect to the Bernoulli distribution $B_\varepsilon$\textup)
subset $E\subset \mathbb{Z}^2$, every $(\tau,E)$-tiling
is at most $1/10$
Besicovitch apart from some $\tau$-tiling of the entire
plane~$\mathbb{Z}^2$;

 \textup{(4)}  for all sufficiently small $\varepsilon$ for almost every
$B_\varepsilon$-random
subset $E\subset \mathbb{Z}^2$, for every $(\tau,E)$-tiling $T$
the Kolmogorov
complexity of centered squares of $T$  of size $N\times N$
is $\Omega(N)$.
\end{thm}
The rest of the section is devoted to the proof of this theorem.
It combines almost all technique developed in this paper:
self-similar tile sets with variable zoom factors, embedding
a sequence  with Levin's property
(i.e., with linear Kolmogorov complexity of all
factors) into tilings, bi-sparse sets, incremental error correcting,
and robustness against doubled holes.

In this section the basic idea of incremental error correcting is applied  in
a slightly modified form. Here we cannot apply
directly the technique of $(c_1,c_2)$-robustness from Section~\ref{robust}.
Instead we use the idea of robustness against holes of some sequence of sizes
$\Delta_0, \Delta_1, \Delta_2, \ldots$, as explained in Section~\ref{robust-var}.
More precisely, we do it as follows: We split the set of random errors into
bi-islands of different ranks. Then, we eliminate them one by one, starting from
lower ranks.
When we correct an isolated bi-island of rank $k$, we need a precondition
(similarly to the argument in Section~\ref{robust}): In
a large enough neighborhood of this bi-island there are no other errors. Elimination of
a $k$-level bi-island involves corrections in its \emph{extended} $O(\Delta_k)$-neighborhood
(with all parameters as specified below).

\subsection{The main difficulties and ways to circumvent them}

We want to combine the construction from Section~\ref{complex}
with error-correcting methods based on the idea of ``islands'' of errors.
There are two main difficulties in this plan: fast growing zoom factors
and gaps in vertical columns. Let us discuss these two problems in some
detail.

The first problem is that our construction of  tiling with high Kolmogorov
complexity from Section~\ref{complex}
requires \emph{variable zoom factors}. What is even worse is that
zoom factors $N_k$  must increase very fast
(with logarithms growing faster than $2^k$).
Hence, we cannot  directly apply the technique of islands from
Section~\ref{islands} since it works only when
$\sum\frac{\log \beta_k}{2^k}<\infty$ (where $\beta_k$ is the parameter from
the definition of islands, which, in our construction,  must be of the same order as
the size of $k$-level macro-tiles).
To overcome this obstacle, we replace islands by bi-islands
(the technique developed in Section~\ref{bi-islands}).

The second problem is that now we cannot reconstruct a macro-tile from
the information ``consciously known''  to this macro-tile.
The missing information is the sequence of bits
assigned to  the vertical columns (with each vertical column of tiles carrying  one bit of a high-complexity sequence $\omega$).
Random errors make gaps in vertical
columns, so now the columns are split into parts, which \textit{a priori} can  carry different bits.
To overcome this problem we organize additional information flows between
macro-tiles   to guarantee that  each infinite vertical column
carries in most  of its tiles one and the same bit value.

\subsection{General scheme}\label{section-13-2}

Here we explain the general ideas of  our proof. First, we use
macro-tiles with variable zoom factors $N_k=Q^{\lfloor 2.5^k \rfloor}$ for a large
enough integer $Q>0$.  This means that every $k$-level macro-tile is an
$(N_{k-1}\times N_{k-1})$ array of $(k-1)$-level macro-tiles. So  the size (the number
of columns and the number of rows) of a $k$-level macro-tile is
$L_k=N_0\cdots N_{k-1}$, and $L_k<N_k$. (The constant $2.5$ in our construction
can be replaced  by any rational number  between $2$ and $3$.)

To get tilings with high Kolmogorov complexity, we reuse the construction
from Section~\ref{complex} with the  zoom factors defined above.
Let us recall the idea of that construction
(proof of Theorem~\ref{thm:complex}). In a correct tiling,
in the $i$th column  all tiles keep some bit
 $\omega_i$, and we want every $N$-bit substring
 in the corresponding biinfinite sequence $\omega$
to have Kolmogorov complexity $\Omega(N)$.
To enforce this property we organize our computation on macro-tiles
of all levels. The crucial point of the construction is
propagation of bits $\omega_i$ to the computation zones
of macro-tiles of high levels.
Let us recall the main points of this construction (following the argument from Section~\ref{complex}):
\begin{itemize}\label{section7-outline}
\item We say that for each (infinite) column of tiles in a tiling there is an \emph{assigned}
bit $\omega_i$, which is ``known'' to each tile in the column.
(In other words, there is a mapping that attributes to each tile the corresponding bit $\omega_i$;
vertically neighboring tiles must keep the same value of the bit.)
\item For a $k$-level macro-tile (of size $L_k\times L_k$) its \emph{zone of responsibility} is the  sequence of $L_k$
bits $\omega_i$ assigned to all columns of this macro-tile. Vertically aligned macro-tiles of the same level have the same zone of responsibility.
\item For \emph{some} $k$-level macro-tile $M$ there is one \emph{delegated bit};  this is a bit $\omega_i$ from the zone of responsibility of this macro-tile. This bit must be known to the ``consciousness'' of the macro-tile; that is, it must be presented
explicitly on the tape in the computation zone of this macro-tile.
For technical reasons, we decide that the position  of the delegated bit $\omega_i$  in the zone of responsibility of $M$  (this position is an integer between $0$ and $L_k-1$) is equal to the position (vertical coordinate) of $M$ in its father macro-tile (see Fig.~\ref{fpt.8.mps}).  The father is a macro-tile of level $k+1$, which consists of $N_k\times N_k$ macro-tiles of level $k$ (thus, the vertical coordinate of a $k$-level macro-tile in its father ranges over $0,\ldots, N_k-1$). In our settings, $N_k > L_{k-1}$. If a $k$-level macro-tile $M$ has a vertical coordinate in its father greater than $N_k$, then $M$ does not have  a delegated bit.
\item If a $k$-level macro-tile $M$ has a delegated bit in its computation zone, it
also contains  a \emph{group of bits to check} that starts at the delegated bit and has
rather small length (say, $\log\log\log k$). If this group of bits leaves the responsibility zone,
we truncate it. The Turing machine simulated in the computation zone of $M$  enumerates the forbidden strings  of
``too small Kolmogorov complexity''  and verifies  that the \emph{checked group of bits} does not contain
any of them. This process is bounded by time and space allocated to the computation zone of
a $k$-level macro-tile.
\end{itemize}
The last item requires additional comments. Technically, we fix constants $\alpha\in(0,1)$ and $c$ and
check that for every string $x$ in  zones of responsibility of all macro-tiles $K(x)\ge\alpha|x|-c$. To check this
property, a macro-tile enumerates all strings $x$ of complexity less than $\alpha|x|-c$. This enumeration requires infinite time, though computations in each macro-tile are time-bounded. However, this is not a problem since every such $x$ is checked in macro-tiles of arbitrarily high levels
(i.e., if $x$  is covered by a macro-tile of level $k$, then it is also covered by macro-tiles of all levels greater than $k$).
Thus, we guarantee the following property:
 \begin{equation}\label{levin-property}\tag{\hbox{$*$}}
 \begin{array}{c}
 \parbox{0.7\textwidth}{\emph{For every $k$-level macro-tile $M$
 \textup($k=1,2,\ldots$\textup), and
for every  substring $x$
of $\omega$ that is contained in $M$'s zone
of responsibility \textup({its horizontal projection}\textup),
it holds that
$K(x) \ge \alpha|x|-c$.}}
 \end{array}
 \end{equation}
Notice that   $K(x) \ge \alpha|x|-c$ holds  only for strings $x$
covered by some macro-tile (i.e., strings that belong to some macro-tile's zone of responsibility). In
``degenerate'' tilings there can exist an infinite
vertical line that is a border line for macro-tiles of \emph{all} levels
(see Fig.~\ref{degenerate.mps}).  A string $x$ that intersects this line
is not covered by any macro-tile of any level. Hence,~($*$) does not guarantee
for such a string $x$ that its Kolmogorov complexity is greater than $\alpha|x|-c$. However, as
we noticed in Section~\ref{last-correction},
the parts of $x$ on both sides of the boundary are covered by some macro-tile.
Hence, it follows from~($*$) that $K(x)\ge \frac{\alpha}{2}|x|-O(1)=\Omega(|x|)$ for \emph{all} factors $x$ of the biinfinite string $\omega$.

Thus, we reuse  the argument from Section~\ref{complex}, and  it
works well if there are no errors, but when we introduce random
errors, the old construction is broken. Indeed,
vertical columns can be  damaged by islands of errors.
Now we need to make an  effort to enforce that copies of
$\omega_i$ consciously kept by different macro-tiles
are coherent  (at least for  macro-tiles that are not  seriously
damaged by local errors). To this end we will use some
checksums, which guarantee that neighbor macro-tiles keep
coherent conscious and subconscious information. We discuss
this topic in the next section.

To deal with random errors we use the technique of
bi-islands (see Section~\ref{bi-islands}).
Our arguments work if diameters of $k$-level
bi-islands are comparable with the size of $k$-level macro-tiles.
Technically, we set $\alpha_k=26 L_{k-1}$ and
$\beta_k=2L_{k}$. (In the following we will see that this choice of $\beta_k$ is important
for the error-correcting procedure; $\alpha_k$ is set to $13\beta_{k-1}$, so that
lemmas on bi-islands can be applied.)
Recall that  $N_k=Q^{\lfloor 2.5^k \rfloor}$ and $L_k=N_0\cdots N_{k-1}$.
Note that  Lemmas~\ref{lemma-bi-island} and~\ref{lemma-bi-island-2} can be used
with these values of parameters $\alpha,\beta$.
We will  also employ Lemma~\ref{lemma-bi-island-3} with $\gamma_k=O(\alpha_k)$.

 \subsection{The new construction of the tile set\label{robust-complex-3}}

We take the construction from Section~\ref{complex} as the starting
point and superimpose some new structures on $k$-level macro-tiles.
We  introduce these supplementary structures in several steps.

\textbf{First step (introducing checksums):}
Every $k$-level macro-tile $M$ (in a correct tiling) consists of an
$N_{k-1}\times N_{k-1}$ array
of $(k-1)$-level macro-tiles; each of these $(k-1)$-level macro-tiles may
keep one  delegated bit. Let us take one horizontal row
(bits assigned to $N_{k-1}$ macro-tiles of level $k-1$)
 in this two-dimensional array of size $N_{k-1}\times N_{k-1}$.
 Denote the corresponding sequence of  bits
by $\eta_1,\ldots,\eta_{N_{k-1}}$. We introduce a sort of \emph{erasure code}
for this string of bits. In other words, we will calculate  some checksums for this sequence.
These  checksums should be suitable to reconstruct all bits $\eta_1,\ldots,\eta_{N_{k-1}}$
\emph{if at most $D$ of these bits are erased} (i.e., if we  know values $\eta_i$ for only
$N_{k-1}-D$ positions); here $D>0$ is a constant (to be fixed later).
We want the checksums to be easily computable. Here we use again the
checksums of the Reed--Solomon code (discussed in
Section~\ref{strongly}).

Let us explain this technique in more detail. We take a finite field
$\mathbb{F}_k$ of large enough size  (greater than $N_{k-1}+D$). Then, we calculate
a polynomial of degree less than $N_{k-1}$
that  takes values $\eta_1,\ldots,\eta_{N_{k-1}}$
at some  $N_{k-1}$ points of the field.
Further, we take as checksums the values of this polynomial
at some other $D$  points from $\mathbb{F}_k$ (where all $(N_{k-1}+D)$
points of the field are fixed in advance).
Two  polynomials of degree less than $N_{k-1}$ can coincide
in at most  $(N_{k-1}-1)$ points. Hence,
if  $D$ bits from the sequence
$\eta_1,\ldots,\eta_{N_{k-1}}$ are erased, we can reconstruct
them given the other (nonerased) bits $\eta_j$ and the checksums defined above.

These checksums contain $O(\log N_{k-1})$ bits of
information. We  next discuss how to compute them.

\textbf{Second step (calculating checksums):}
First, we explain how to compute the checksums,
going from left to  right along the sequence
 $\eta_1,\ldots,\eta_{N_{k-1}}$.
This can be done in a rather standard way as follows.

Let $\eta_1,\ldots,\eta_{N_{k-1}}$ be the values of a polynomial
$p(x)$ (of degree less than $N_{k-1}$)
at points $x_1,\ldots,x_{N_{k-1}}$. Assume we want
to reconstruct all coefficients of this polynomial. We can
do this by the following iterative procedure. For $i=1,\ldots,N_{k-1}$
we calculate polynomials
 $ p_i(x) $ and $q_i(x)$ (of degree $\le (i-1)$ and $i$, respectively)
 such that
$$ p_i(x_j) = \eta_j\ \mbox{ for }\ j=1,\ldots,i $$
and
 $$q_i(x) = (x-x_1)\cdots(x-x_i).$$
It is easy to see that  for each $i$, polynomials $p_{i+1}$ and $q_{i+1}$
can be computed from polynomials $p_{i}$ and $q_{i}$
and the values $x_{i+1}$ and $\eta_{i+1}$.

If we do not need to know the resulting polynomial $p=p_{N_{k-1}}(x)$
but want to get only the value $p(a)$ at some particular point $a$,
then we can perform all these calculations modulo $(x-a)$.
Thus, to obtain the value of $p(x)$ at $D$ different points,
we run in parallel $D$ copies of this process.
At each step of the computation  we need to keep in memory only
$O(1)$ elements of $\mathbb{F}_k$, which is
$O(\log N_{k-1})$ bits of temporary data
(with the multiplicative constant in this $O(\cdot)$ notation depending
on the value of $D$).

This calculation can be  simulated by a tiling. We embed
the procedure just explained into the computation zones of
$(k-1)$-level macro-tiles. The partial results of the calculation are transferred
from one $(k-1)$-level macro-tile to another one, from the left to the right
(in each row of length $N_{k-1}$ in a $k$-level macro-tile). The final result
(for each row) is embedded into the conscious information (bits on the
tape of the Turing machine in the computation zone) of the
rightmost $(k-1)$-level macro-tile of the row.

To organize these computations,
we need to  include into conscious information kept by  $(k-1)$-level
macro-tiles additional $O(\log N_{k-1})$ bits and add the same number of bits
to their macro-colors. This fits well
our fixed-point construction since zoom factors $N_k$ grow fast,
and we have enough room in the computation zone.

\begin{figure}[h]
\begin{center}
$$
\includegraphics[scale=0.8]{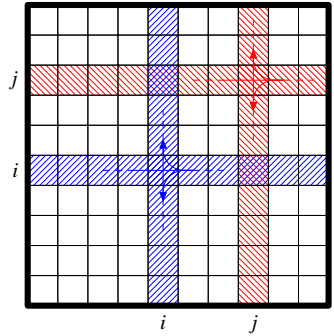}
$$
\end{center}
\caption{Propagation of checksums inside of a macro-tile.}
\label{checksum}
\end{figure}

\textbf{Third step (consistency of checksums between macro-tiles):}
So far, every $k$-level macro-tile  contains
$O(N_{k-1}\log N_{k-1})$ bits of checksums and $O(\log N_{k-1})$ bits
for every row.
We want these checksums to be the same for every two
vertical neighbor macro-tiles. It is inconvenient  to keep
the checksums for all rows only in the rightmost column (since
it would create too much traffic in this column if we try
to transmit the checksums to the neighbor macro-tiles of level $k$).
So we propagate the checksums of the $i$th row in a $k$-level
macro-tile $M$ ($i=1,\ldots, N_{k-1}$)
along the entire $i$th row and  along the entire $i$th column
of $M$.  In other words, these checksums must be ``consciously'' known
to all  $(k-1)$-level macro-tiles in the $i$th row and in the $i$th  column
of $M$. In Fig.~\ref{checksum}
we show the area of propagation of checksums for two rows
(the $i$th  and the $j$th rows).

On the border of two neighbor $k$-level macro-tiles (one above
another) we check that in each column $i=1,\ldots, N_{k-1}$
all the corresponding checksums computed in both macro-tiles
coincide. This check is redundant if there are no errors in the tiling:
The checksums are computed from the delegated bits (which come
from the sequence of bits $\omega$ encoded into tiles
of the ground level), so the corresponding values
for all vertically aligned macro-tiles must be equal to each other.
However, this redundancy is useful to resist  errors,
as we show in the following.

\textbf{Fourth step (robustification):}
The  features just explained organized in every $k$-level macro-tile
(bit delegation, computing and propagating
checksums, and all the computations simulated in the computation zone
of a macro-tile) are  simulated by means of bits kept in the ``consciousness''
(i.e., in the computation zone)
of $(k-1)$-level macro-tiles. Now we fix some constant $C$ and
``robustify''
this construction in the following sense: Each $(k-1)$-level macro-tile $M$
keeps in its consciousness not only ``its own'' data but also the bits
previously assigned to $(k-1)$-level macro-tiles from  its
$(C\cdot L_{k-1})$-neighborhood  (i.e.,  the $(2C+1)\times (2C+1)$
array of $(k-1)$-level macro-tiles centered at $M$). So,
the content of the consciousness of each macro-tile is multiplied
by some constant factor. Neighbor macro-tiles check that the data in their
consciousness are coherent.

We choose the constant $C$ so that every $k$-level bi-island (which consist of two
parts of size $\alpha_k$)
and even the $\gamma_k=O(\alpha_k)$-neighborhood of every
$k$-level bi-island (where we specify $\gamma_k$ below)
can involve only a small part of the $(CL_{k-1})$-neighborhood
of any $(k-1)$-level macro-tile. (Note that here we talk about neighborhoods, not
about \emph{extended neighborhoods} of bi-islands defined in Section~\ref{bi-islands}.)

This robustification allows us to reconstruct the conscious information of
a $k$-level macro-tile and of its  $(k-1)$-level sons when this macro-tile is damaged
by one $k$-level bi-island (assuming there are no other errors).

\textbf{The last remark (the number of  bits in the consciousness of a macro-tile):}
The construction explained above requires that
we put into the computation zones of  all $(k-1)$-level
macro-tiles additional $\poly(\log N_{k-1})$ bits of data.
(The most substantial part of the data is the information used to compute
the checksums.)
Again, this fits  our fixed-point construction because
$\poly(\log N_{k-1})$ is much less than $N_{k-2}$, so
we have enough room to keep and process all these data.

\medskip

The tile set $\tau$ is thus defined. Since there exists an $\omega$ with Levin's
property, it follows that $\tau$-tiling exists, and every $N\times N$ square
of such a tiling has Kolmogorov complexity $\Omega(N)$. Further,
we prove that this $\tau$ satisfies also statement (3) of
Theorem~\ref{thm-complex-errors}.

\subsection{Error-correcting procedure\label{robust-complex-4}}

Denote by $\tau$ the tile set described in Section~\ref{robust-complex-3}.
Let $\varepsilon>0$ be small enough. Lemma~\ref{biisland}
says that a $B_\varepsilon$-random set with probability $1$
is bi-sparse. Now we assume that $E\subset \mathbb{Z}^2$ is a
bi-sparse set (for the chosen values of $\alpha_i$ and $\beta_i$), and $T$ is a $\tau$-tiling
of $\mathbb{Z}^2 \setminus E$. Further, we explain how to correct
errors and convert $T$ into a tiling $T'$ of the entire plane (where $T'$ should be
close to $T$).

We follow the usual  strategy. The set $E$ is bi-sparse; that is,
it can be represented as a union of isolated bi-islands of different ranks.
We correct them one by one, starting from bi-islands of low ranks. To prove
that the correction procedure converges, we
need  to explain one step of this process:
how to correct one bi-island $S$ of rank $k$ assuming
that  it is well isolated, i.e., in the  $\beta_k$-neighborhood of this bi-island
there are no other (still noncorrected) errors.

Let us recall that a $k$-level bi-island $S$ is a union of two ``clusters''
$S_0, S_1$; the diameters of both $S_0$ and $S_1$
are at most $\alpha_k =O( L_{k-1})$.
Hence the clusters $S_0$ and $S_1$ touch only $O(1)$ macro-tiles
of level $(k-1)$.
The distance between $S_0$ and $S_1$ is at most $\beta_k$,
and the $\beta_k$-neighborhood of $S$ is free of other bi-islands of rank
$k$ and higher (so we can assume that  the $\beta_k$-neighborhood
of $S$ is already cleaned of errors).
Our correction procedure around $S$
will involve only points in the extended $\gamma_k$-neighborhood of~$S$,
where $\gamma_k=2 \alpha_k$.

Let $M$ be one of $k$-level
macro-tiles intersecting the extended $\gamma_k$-neighborhood
of the $k$-level bi-island $S$. Basically, we need to reconstruct
all $(k-1)$-level macro-tiles in $M$ destroyed by $S$. First, we will reconstruct
the  conscious information in all $(k-1)$-level macro-tiles in $M$.
This is enough to get all bits of $\omega$
from the ``zone of responsibility'' of $M$. Then, we will reconstruct  in a consistent way
all $n$-level macro-tiles inside $M$ for all $n<k$.

Thus, we start with reconstructing the consciousness of all $(k-1)$-level
macro-tiles $M'$  in $M$.
First, we recall that the consciousness (the content of the computation
zone) of every $(k-1)$-level macro-tile $M'$ consists of several groups of bits
(cf. the outline of the construction in Section~\ref{section-13-2},
p.~\pageref{section7-outline}):
 \begin{enumerate}
 \item[\hbox{[A]}] the binary representation of the number $(k-1)$ and
 coordinates (integers from the range $0,\ldots, N_{k-1}-1$) of $M'$ in the father macro-tile $M$;
  \item[\hbox{[B]}] the bits used to  simulate a Turing machine on the computation
 zone of  $M$ and the bits used to implement ``wires'' of  $M$;
 \item[\hbox{[C]}] the bit (from the sequence $\omega$) delegated to
 $M'$;
 \item[\hbox{[D]}] the bit (from  $\omega$) delegated to $M$;
 \item[\hbox{[E]}] the bits used to calculate and communicate the checksums
 for the corresponding row of $(k-1)$-level macro-tiles  in  $M$;
and \item[{[F]}]  a group of bits to check  from the zone of responsibility of $M'$; these bits are checked by the macro-tile: $M'$ checks on its computation zone that this ``group of bits to check'' does not contain any factor of low Kolmogorov complexity.
 \end{enumerate}
Bits of field [A] in a small isolated group of $(k-1)$-level macro-tiles
are trivially reconstructed from the surrounding macro-tiles of the same level.
Fields [B], [C], [D], and [E] can be reconstructed
because of the robustification on the level of $(k-1)$-level macro-tiles.
(We organized the robustification on the level of $(k-1)$-level macro-tiles
 in such a way that we are able to reconstruct these fields
for any $C\times C$ group of missing or corrupt
$(k-1)$-level macro-tiles.)
So far the correcting  procedure follows the exactly the same steps as in
Section~\ref{robust-var}.

To reconstruct fields [F] of $(k-1)$-level macro-tiles in $M$, we need to
reconstruct all bits of $\omega$ from the zone of responsibility of $M$.
We can extract these bits from the neighbor $k$-level tiles above or below $M.$
(Recall that bi-island $S$ touches only $O(1)$ $k$-level
macro-tiles, and there is a ``healthy'' zone of $k$-level macro-tiles around them.)
However, a problem remains since we are not sure that  the $\omega$ bits
above $M$, below $M$, and inside $M$ are consistent.  Now we show that this
consistency is guaranteed by checksums.

\begin{figure}[h]
\begin{center}
\includegraphics[scale=0.6]{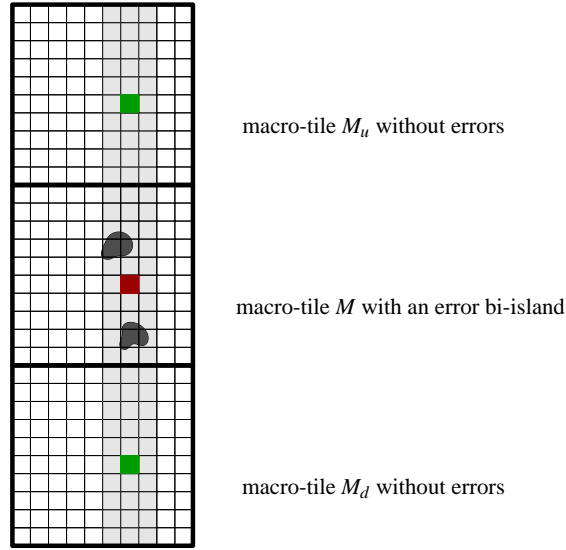}\hspace*{2cm}
\end{center}
\caption{Bi-island of errors in a macro-tile.}
\label{biisland-correcting}
\end{figure}

Denote by $M_u$ and $M_d$ the $k$-level
macro-tiles just above and below $S$. Since the distance between $S$ and
other $k$-level bi-islands is greater than $\beta_k=2L_k$, we know that $M_u$ and $M_d$
must be free of errors (where we assume that errors of ranks less than $k$ are already corrected).
See Fig.~\ref{biisland-correcting}.  In what follows,
our explanations refer to Fig.~\ref{biisland-correcting},
where bi-island $S$ touches  only one $k$-level macro-tile; if $S$ touches
several $k$-level macro-tiles, substantially the same arguments work.
It is enough to prove that the  bits $\omega_i$ assigned to corresponding columns of
$M_u$ and in $M_d$ are equal to each other.

The macro-tiles $M_u$ and $M_d$ are error free; therefore, the
 sequences of $L_k$ bits $\omega_i$ corresponding to the vertical lines
 intersecting these $k$-level macro-tiles are well defined.
 Since there are no errors, the conscious information (including checksums)
in all macro-tiles of all  levels inside $M_u$ and $M_d$
is consistent with these bit sequences. So, the $L_k$ bits assigned to the vertical
columns are correctly delegated to the corresponding $(k-1)$-level macro-tiles
inside $M_u$ and $M_d$. However, it is not evident that the sequences
of $L_k$ bits embedded in $M_u$ and $M_d$ are equal to each other.

In fact, it is easy to see that bit sequences for $M_u$ and $M_d$ coincide with each other
at most positions. They must be equal for all columns (from the range $0,\ldots, L_k-1$)
that do not intersect bi-island $S$ (i.e., in nondamaged columns of tiles on the ground level,
the assigned bits $\omega_i$ correctly spread though macro-tiles $M_u$, $M$, and $M_d$).
Hence, the bits delegated to the corresponding $(k-1)$-level
macro-tiles in $M_u$ and $M_d$ are equal to each other, except for only $(k-1)$-level
macro-tiles in the ``gray zone''  of Fig.~\ref{biisland-correcting}, which contains
the $(k-1)$-level macro-tiles involved in the correction of $S$ and all vertical stripes
touching the involved sites.
(The width of this gray stripe is only $O(1)$ macro-tiles of level $(k-1)$.) Hence, for $i=0,\ldots, (N_{k-1}-1)$,
in the $i$th rows of $(k-1)$-level macro-tiles
in $M_u$ and $M_d$,
the sequences of delegated bits are equal to each other except possibly for only $O(1)$ bits
(delegated to $(k-1)$-level macro-tiles in the ``gray zone'').

The robustness property guarantees that  all checksums are correctly transmitted
through $M$. Hence, checksums for corresponding rows in $M_u$ and in $M_d$
must be  equal to each other.

Thus, for every two corresponding rows of $(k-1)$-level macro-tiles  in $M_u$ and in $M_d$
we  know that (a) all except  $O(1)$ delegated bits in the corresponding positions
are equal to each other and (b) the checksums  are equal to each other. From the property
of our erasure code it follows that in fact \emph{all} delegated bits in these rows
are equal to each other (with every $i$th bit in $M_u$ being equal to the $i$th bit in $M_d$).
Therefore, all bits $\omega_i$ in $M_u$ and $M_d$
are the same (on the ground level). We can use these bits to reconstruct subconsciousness
of $M$ and get a consistent tiling in $M$.

We are almost done. Bi-island $S$ is corrected; we reconstructed conscious information for the
$k$-level macro-tile $M$ and for all its $(k-1)$-level sons.
Now we can reconstruct fields [F] in the damaged $(k-1)$-level macro-tiles inside $M$.
This is simple to do. We just take the corresponding bits $\omega_i$ from the zone of responsibility
(shared by $M$, $M_u$, and $M_d$).
It remains only to explain why the checking procedure does not fail for these groups of bits
(i.e., $(k-1)$-level macro-tiles do not discover in these bit strings any factors of low Kolmogorov complexity).
But this is true because  macro-tiles of levels $(k-1)$ (and also below $(k-1)$)
inside $M$ apply exactly all the same checks to exactly the same groups of bits $\omega_i$ as
the  macro-tiles in the corresponding positions in $M_u$ and $M_d$.
Since there is no errors in $M_u$ and $M_d$,  these computations do not lead to a contradiction.

Let us inspect again the correction procedure just explained; we should notice
which tiles are involved in the error-correcting process around bi-island $S$.
In the $(k-1)$-level macro-tiles
outside the ``gray zone'' we change nothing. Moreover, not all the gray zone needs to be changed---only the part between two clusters of $S$ (and their small neighborhoods) is affected.
Indeed, in all tiles of $M$ that are above $S$ the assigned bits $\omega_i$ are the same as in
the corresponding columns of $M_u$; in the tiles of $M$ that are below $S$ the assigned bits
$\omega_i$ are  the same as in the corresponding columns of $M_d$. Hence, there is no need
to correct ``subconscious information'' of $(k-1)$-level macro-tiles that are above or below $S$.
Only the area between two clusters of $S$ requires corrections.
More precisely, the area involved in the correcting procedure is inside the extended neighborhood
of $S$. (In fact, this argument is the motivation of our definition of extended neighborhood.)

Thus, we have proven that this step-by-step correcting procedure eliminates all bi-islands
of errors and only extended $\gamma_k$-neighborhoods of $k$-level bi-islands
are involved in this process. Now Theorem~\ref{thm-complex-errors}~(part 3)
follows from Lemma~\ref{lemma-bi-island-3}. It remains only to prove
part~4 of the theorem. We do this in the next section.

\subsection{Levin's property for $\omega$ embedded into
a $(\tau,E)$-tiling \label{robust-complex-5}}

It remains to prove part (4) of Theorem~\ref{thm-complex-errors}.
In the previous section we proved that if the set of errors $E$ is bi-sparse, then
a $(\tau,E)$-tiling $T$ can be converted into a $\tau$-tiling $T'$ of the entire plane,
and the difference between $T$ and $T'$ is covered by extended $\gamma_k$ neighbors
of $k$-level bi-islands from $E$ ($k=0,1,\ldots$). Now we want to show that,  in the initial
tiling $T$, the Kolmogorov complexity of centered squares of size $N\times N$ was 
$\Omega(N)$.

Fix a point $O$. Since $E$ is bi-sparse, $O$ is covered
by $\beta_k$-neighborhoods of only finitely many bi-islands. Hence,
for large enough $\Delta$, the $\Delta\times\Delta$ square
$Q_\Delta$ centered at $O$ intersects extended
$\gamma_k$-neighborhoods of $k$-level bi-islands only if
$\beta_k<\Delta$. (If the extended $\gamma_k$-neighborhood of some bi-island
intersects $Q_\Delta$ and $\beta_k\ge \Delta$, then $\beta_k-\gamma_k>\Delta/2$
and $O$ is covered by the $\beta_k$-neighborhood of this bi-island.)
Therefore, to reconstruct $T'$ in $Q_\Delta$ it is enough to correct there all
bi-islands of bounded levels (such that $\beta_k<\Delta$).

To reconstruct $T'$ in $Q_\Delta$ we need to know the original tiling $T$ in  $Q_\Delta$ and some neighborhood
around it  (i.e., in some centered $O(\Delta)\times O(\Delta)$ square $Q_{\Delta'}$, which is only
greater than $Q_\Delta$ by a constant factor). Indeed, given the tiling $T$ restricted on $Q_{\Delta'}$, we can locally correct
there bi-islands of levels $1, 2, \ldots, k$ (such that $\beta_k<\Delta$) one by one.
Correcting a bi-island of errors in $Q_{\Delta'}$ we obtain the same results as in the error-correcting procedure
on the entire plane $\mathbb{Z}^2$ unless this bi-island is too close to the border  of $Q_{\Delta'}$
(and the local correction procedure should involve  information outside $Q_{\Delta'}$).
Thus, we can reconstruct $T'$-tiling not in the entire $Q_{\Delta'}$
but in  points that are far enough from the border
of  this square.
If $\Delta'= c\Delta$ for large enough $c$, then $Q_{\Delta'}$ provides enough information to reconstruct
$T'$ in $Q_\Delta$.

We know that Kolmogorov complexity of error-free tiling $T'$ in $Q_{\Delta}$ is $\Omega(\Delta)$.
Therefore, the Kolmogorov complexity of the original $T$-tiling in
the greater square $Q_{\Delta'}$ is also $\Omega(\Delta)$.
Since $\Delta'$ is only greater than $\Delta$ by a constant factor,
we get that the Kolmogorov complexity of  the $(\tau,E)$-tiling $T$ restricted
to  the centered $(\Delta'\times \Delta')$ square is $\Omega(\Delta')$.

Theorem~\ref{thm-complex-errors} is proven.

\section*{Acknowledgments}

The results included in this paper were discussed with many colleagues, including
Peter G\'acs, Leonid Levin, and our french collaborators Meghyn Bienvenu, 
Laurent Bienvenu, Emmanuel Jeandel, Gregory Lafitte, Nicolas Ollinger, and
Michael Weiss. We are grateful to all of them and to the participants of seminars (the Kolmogorov
seminar at Moscow State University, the IITP seminars, and the FRAC seminar) and conferences (DLT~2008 and ICALP~2009)
where some of these results were presented.
We appreciate the  detailed comments and many useful suggestions given by  the anonymous referee.


\begin{thebibliography}{34}

\bibitem{durand-gurevich}
C.~Allauzen, B.~Durand, Appendix A: Tiling Problems, in
E.~B\"orger, E.~Gr\"adel, Y.~Gurevich, \emph{The Classical
Decision Problems}, Springer-Verlag, Berlin, 1996.

\bibitem{sablik-aubrun}
N.~Aubrun, M.~Sablik, Simulation of recursively enumerable subshifts by
two-di\-men\-si\-o\-nal SFT and a generalization, preprint available at the
home page of M.~Sablik, \newline
\texttt{http://www.latp.univ-mrs.fr/\~\relax sablik/article/SimulSRE.pdf} (as of August 5, 2010).

\bibitem{berger}
R.~Berger, The Undecidability of the Domino Problem,
\emph{Mem. Am. Math. Soc.},
\textbf{66}, 1--72, 1966.

\bibitem{code-book} E.~R.~Berlekamp, \emph{Algebraic Coding Theory}, Aegean Park, Laguna Hills, CA, 1984.


\bibitem{jac2008}
L.~Bienvenu, A.~Romashchenko, A.~Shen, Sparse Sets,
\emph{Journ\'ees Automates Cellulaires 2008 \textup(Uz\`es\textup)},
18--28, Moscow Center for Continuous Mathematical Education, Moscow, 2008, available online at
http://hal.archives-ouvertes.fr/docs/00/27/40/10/PDF/18-28.pdf


\bibitem{culik}
K.~Culik, An Aperiodic Set of $13$ Wang Tiles,
\emph{Discrete Math.}, \textbf{160}, 245--251, 1996.

\bibitem{dls}
B.~Durand, L.~Levin, A.~Shen, Complex Tilings,
\emph{J. Symbolic Logic}, \textbf{73}(2), 593--613, 2008;
see also \textit{Proc.} \emph{33rd Ann. ACM Symp. Theory  Computing},
pp. 732--739, 2001,
and \texttt{www.arxiv.org/cs.CC/0107008} for an earlier version.

\bibitem{intelligencer}
B.~Durand, L.~Levin, A.~Shen, Local Rules and Global Order, or
Aperiodic Tilings, \emph{Math. Intelligencer},
\textbf{27}(1), 64--68, 2004.

\bibitem{durand-romash} B.~Durand, A.~Romashchenko, On
Stability of Computations by Cellular Automata, in \textit{Proc.}
\emph{European Conf. Compl. Syst.}, Paris, 2005.

\bibitem{dlt}
B.~Durand, A.~Romashchenko, A.~Shen,
Fixed Point and Aperiodic Tilings, in
\emph{Developments in Language Theory, 12th International
Conference, DLT 2008, Kyoto, Japan, September 16--19, 2008,
Proceedings}, Lecture Notes in Computer Science, \textbf{5257},
Springer-Verlag, Berlin, 276--288, 2008.

\bibitem{icalp}
B.~Durand, A.~Romashchenko, A.~Shen,
High Complexity Tilings with Sparse Errors, in
\emph{Automata, Languages and Programming, 36th International Colloquium,
ICALP 2009, Rhodes, Greece, July 5--12, 2009, Proceedings, Part I,} Lecture
Notes in Computer Science, \textbf{5555}, Springer-Verlag, Berlin, 403--414, 2009.

\bibitem{gacs-focs} P.~G\'acs, Reliable Cellular Automata with
Self-Organization, in \textit{Proc}.
\emph{38th Annu. Symp. Found. Comput. Sci.},  90--97, 1997.

\bibitem{gacs} P.~G\'acs, Reliable Cellular Automata with
Self-Organization, \emph{J. Stat. Phys.},
\textbf{103}(1/2), 45--267, 2001.

\bibitem{gray} L.~Gray, A Reader's Guide to G\'acs' Positive Rates
Paper, \emph{J. Stat. Phys.},
\textbf{103}(1/2), 1--44, 2001.

\bibitem{grunbaum}
B.~Gr\"unbaum, G.C.~Shephard, \emph{Tilings and Patterns},
Freeman, New York, 1987.

\bibitem{gurevich-koryakov}
Yu.~Gurevich, I.~Koryakov, Remarks of Berger's paper on the domino
problem, \emph{Siberian Math. J.}, \textbf{13}, 319--321, 1972.

\bibitem{hanf}
W.~Hanf, Nonrecursive Tilings of the Plane, I, \emph{J. Symbolic
Logic}, \textbf{39}, 283--285, 1974.

\bibitem{hochman}
M.~Hochman, On the Dynamic and Recursive Properties of Multidimensional
Symbolic Systems, \emph{Inventiones Math.}, \textbf{176}, 131--167, 2009.


\bibitem{kari}
J.~Kari, A Small Aperiodic Set of Wang tiles,
\emph{Discrete Math.},
\textbf{160}, 259--264, 1996.


\bibitem{kleene}
H.~Rogers, \emph{The Theory of Recursive Functions and Effective Computability},
Cambridge, MIT Press, 1987.

\bibitem{lafitte-weiss}
G.~Lafitte and M.~Weiss, Computability of Tilings,
in \textit{Proc.} \emph{International Federation for Information Processing,
Fifth IFIP International Conference on Theoretical Computer Science} (IFIP-TCS 2008),
Vol.~273, 187--201, 2008.

\bibitem{levin-arxiv}
L.~Levin, Aperiodic Tilings: Breaking Translational Symmetry,
\emph{Comput. J.}, \textbf{48}(6), 642--645, 2005, available
online at \texttt{http://www.arxiv.org/cs.DM/0409024}.

\bibitem{finite-fields} R.~Lidl and H.~Niederreiter, \emph{Finite Fields}, 2nd ed., Cambridge University Press, Cambridge, 1997.

\bibitem{mozes}
S.~Mozes,
Tilings, Substitution Systems and Dynamical Systems Generated by
Them, \emph{J. Analyse Math.}, \textbf{53}, 139--186, 1989.

\bibitem{myers}
D.~Myers, Nonrecursive Tilings of the Plane, II, \emph{J. Symbolic
Logic}, \textbf{39}, 286--294, 1974.

\bibitem{neumann} J.~von~Neumann, \emph{Theory of
Self-reproducing Automata}, edited by A.~Burks, University of
Illinois Press, Champaign, IL, 1966.

\bibitem{ollinger} N.~Ollinger,
Two-by-two Substitution Systems and the Undecidability
of the Domino Problem, in \textit{Proc.} \emph{Computability in Europe},
LNCS \textbf{5028}, 476--485, 2008.

\bibitem{pri-ul}
Yu.~Pritykin, J.~Ulyashkina, Aperiodicity Measure for Infinite Sequences,
Computer Science---Theory and Applications, in \textit{Fourth International
Computer Science Symposium in Russia, CSR 2009, Novosibirsk,
Russia, August~18--23, 2009}, Lecture Notes in Computer
Science, \textbf{5675}, Springer-Verlag, Berlin, 2009, pp.~274--285.

\bibitem{robinson}
R.~Robinson, Undecidability and Nonperiodicity for Tilings of
the Plane, \emph{Inventiones Mathem.},
\textbf{12}, 177--209, 1971.

\bibitem{rumush}
An.~Rumyantsev, M.~Ushakov, Forbidden Substrings, Kolmogorov
Complexity and Almost Periodic Sequences, in \emph{STACS 2006
Proceedings}, Lecture Notes in Computer Science, \textbf{3884},
Springer-Verlag, Berlin, 2006.

\bibitem{uppsala-notes}
A.~Shen,
Algorithmic Information Theory and Kolmogorov Complexity,
 lecture notes of a course taught at Uppsala University,  available as a technical report at
 \newline
 \texttt{http://www.it.uu.se/research/publications/reports/2000-034/}.

\bibitem{simpson-subshifts}
S.~G.~Simpson, Medvedev degrees of 2-dimensional subshifts of finite type, \emph{Ergodic Theory and Dynamical Systems}, \textbf{34}, 665--674, 2014.


\bibitem{thue-morse-analysis}
M.~Zaks, A.S.~Pikovsky, J.~Kurths, On the Correlation Dimension
of the Spectral Measure for the Thue--Morse Sequence,
\emph{J. Stat. Phys.},
\textbf{88}(5/6), 1387--1392, 1997.

\bibitem{dict}  Merriam-Webster's Medical Dictionary,
 \texttt{http://dictionary.reference.com,} accessed August 10, 2010.



\end{thebibliography}
\end{document}